\documentclass[lettersize,journal]{IEEEtran}
\usepackage{amsmath,amsfonts}
\usepackage{algorithmic}
\usepackage{algorithm}
\usepackage{array}
\usepackage[caption=false,font=normalsize,labelfont=sf,textfont=sf]{subfig}
\usepackage{textcomp}
\usepackage{stfloats}
\usepackage{url}
\usepackage{verbatim}
\usepackage{graphicx}
\usepackage{cite}
\usepackage{multirow}
\usepackage{hyperref}
\hypersetup{hypertex=true}

\makeatletter 
\newcommand\figcaption{\def\@captype{figure}\caption} 
\newcommand\tabcaption{\def\@captype{table}\caption} 
\makeatother

\newcommand{\etal}{\textit{et al.}}
\newcommand{\eg}{\textit{e.g.,}}
\newcommand{\ie}{\textit{i.e.}}

\newcommand{\Tref}[1]{Table~\ref{#1}}

\newcommand{\Fref}[1]{Figure~\ref{#1}}

\usepackage{color}
\newcommand{\hl}[1]{\textcolor{blue}{#1}}
\newcommand{\wc}[1]{\textcolor[RGB]{139,0,0}{#1}} 
\newcommand{\reshl}[3]{{#1}\fontsize{6.5pt}{0.25em}\selectfont{~\hl{(${#2}$\textbf{#3})}}}

\newcommand{\zj}[1]{\textcolor{black}{#1}}

\newcommand{\zjj}[1]{\textcolor{black}{#1}}

\begin{document}
	
	\title{Poison Ink: Robust and Invisible Backdoor Attack}
	
	\author{Jie Zhang, Dongdong Chen$\dagger$, Qidong Huang, Jing Liao, Weiming Zhang, Huamin Feng,\\
		Gang Hua,~\IEEEmembership{Fellow,~IEEE}, and Nenghai Yu,
		\IEEEcompsocitemizethanks{
			\IEEEcompsocthanksitem Jie Zhang, Qidong Huang, Weiming Zhang and Nenghai Yu are with School of Cyber Science and  Security, University of Science and Technology of China, Hefei, Anhui 230026, China. E-mail: \{zjzac@mail., hqd0037@mail., zhangwm@, ynh@\}ustc.edu.cn
			\IEEEcompsocthanksitem Dongdong Chen is with Microsoft Research, Redmond, Washington 98052, USA. E-mail: cddlyf@gmail.com
			\IEEEcompsocthanksitem Jing Liao is with Department of Computer Science, City University of Hong, E-mail: jingliao@cityu.edu.hk
			\IEEEcompsocthanksitem Huaming Feng is with Beijing Electronic Science and Technology Institute, E-mail: fenghm@besti.edu.cn
			\IEEEcompsocthanksitem Gang Hua is with Wormpex AI Research LLC, WA 98004, US, E-mail: ganghua@gmail.com
			\IEEEcompsocthanksitem  $\dagger$ Dongdong Chen is the corresponding author.
		}
	}
	
	\markboth{Journal of \LaTeX\ Class Files,~Vol.~14, No.~8, August~2021}%
	{Shell \MakeLowercase{\textit{et al.}}: A Sample Article Using IEEEtran.cls for IEEE Journals}
	
	
	\maketitle
	
	\begin{abstract}
		Recent research shows deep neural networks are vulnerable to different types of attacks, such as adversarial attacks, data poisoning attacks, and backdoor attacks. Among them, backdoor attacks are the most cunning and can occur in almost every stage of the deep learning pipeline. 
		Backdoor attacks have attracted lots of interest from both academia and industry. However, most existing backdoor attack methods are visible or fragile to some effortless pre-processing such as common data transformations. To address these limitations, we propose a robust and invisible backdoor attack called ``Poison Ink''. Concretely, we first leverage the image structures as target poisoning areas and fill them with poison ink (information) to generate the trigger pattern. As the image structure can keep its semantic meaning during the data transformation, such a trigger pattern is inherently robust to data transformations. Then we leverage a deep injection network to embed such \zj{input-aware} trigger pattern into the cover image to achieve stealthiness. Compared to existing popular backdoor attack methods, Poison Ink outperforms both in stealthiness and robustness. Through extensive experiments, we demonstrate that Poison Ink is not only general to different datasets and network architectures but also flexible for different attack scenarios. Besides, it also has very strong resistance against many state-of-the-art defense techniques.
	\end{abstract}
	
	
	\section{Introduction}
	\IEEEPARstart{I}{n} the past years,  deep learning has achieved tremendous success in a lot of application areas \cite{krizhevsky2012imagenet,graves2013speech,chan2015pcanet, elad2017style, vaswani2017attention}. However, recent works show that they are vulnerable to different types of attacks, such as adversarial attacks \cite{szegedy2013intriguing, dong2018boosting,mustafa2019image,wang2021psat, che2021adversarial}, data poisoning attacks \cite{jagielski2018manipulating, shafahi2018poison} and backdoor attacks \cite{gu2019badnets, liu2017trojaning, chen2017targeted}. Adversarial attacks focus on misleading the model only in the test process, while data poisoning attacks aim to degrade the model inference performance of its primary task by contaminating the training process.
	Backdoor attacks are more flexible and cunning than the two attacks above. 
	Specifically, backdoor attack can affect even all stages of machine learning pipeline \cite{gao2020backdoor}, such as model training \cite{gu2019badnets}, fine-tuning \cite{liu2017trojaning} and even after deployment \cite{rakin2020tbt}. 
	In addition, backdoor attacks are designed to make the backdoored model behave like a normal model unless the attackers feed some specially designed triggers.

	Since it was introduced, an increasing amount of research has been paying attention to this field. For example, Gu \etal \cite{gu2019badnets} directly stamped a square sticker or flower patch onto clean images to contaminate the training process, and Chen \etal \cite{chen2017targeted} replaced the injection strategy by blending a trigger image with the clean example.
	However, there are two main limitations to existing backdoor attack methods. First, the trigger pattern used in many current works is visible (\zj{some visual examples are shown in \Fref{fig:vis_cmp}}), which can be easily recognized by humans or some deep visualization methods
	like Grad-CAM \cite{selvaraju2017grad}. 
	Second, a very recent research \cite{li2020rethinking} finds that most existing backdoor attacks will totally fail during the inference stage if pre-processing the trigger images with some simple data transformations, such as flipping and padding after shrinking.

	To address the above limitations, we aim to design a robust and invisible backdoor attack method.
	Besides the essential requirement for backdoor attacks: 1) the trigger pattern should be easy for the model to learn and cannot confuse the model to affect the pristine performance, we add two more requirements to achieve our goal: 2) the trigger pattern should be consistent under conventional data transformations to keep its robustness; 3) the trigger image needs be visually indistinguishable from the corresponding trigger-free image.
	
	In this paper, we propose to utilize the image structure (edge) of an image as the carrier of poison information, \ie, hiding the poison information into the edge structures as shown in \Fref{fig:fm}. Compared to existing trigger patterns, the proposed structure-based trigger pattern has several natural advantages: 1) On one hand, the shallow layers of DNN  \cite{zeiler2014visualizing} often capture the low-level structure information, which means the structure can be easily captured by DNN; on the other hand, the final decision of DNN \cite{geirhos2018imagenet} often depends on the object texture rather than the structure information, which indirectly indicates that structure-based trigger pattern will not undermine performance of the original task. So it satisfies the first requirement. 2) It is distributed in the whole image and can keep its semantic meaning unchanged under common data transformations, which indirectly satisfies the second requirement. 3) Edge structures belong to high-frequency components of an image, so hiding information into them is more difficult to be discovered, which satisfies the third requirement.

	Based on the above observations, a new backdoor attack method \textbf{``Poison Ink"} is designed. The overall framework is shown in \Fref{fig:fm}. Specifically, we outline the image structures as poison areas and then embed color values ( representation of poison information) into such areas to generate the trigger pattern. To achieve stealthiness, we use a deep injection network to hide the \zj{input-aware} trigger pattern into the cover image to produce the final poisoned image.
	Such poisoned images will be regarded as the trigger set and mixed with clean images as the training set for the standard backdoor training.

	To demonstrate the effectiveness of Poison Ink,  we conduct extensive experiments on various datasets and network architectures under different attacking scenarios.
	Compared to existing backdoor attack methods, Poison Ink outperforms in terms of stealthiness, robustness, generality, and flexibility. In summary, our contributions are four-fold as follows:
	\begin{itemize}
		\item We are the first to proposes utilizing image structures as the carrier of trigger patterns, 
		showing they have natural advantages over existing trigger pattern designs.
		\item We design a new backdoor attack framework, Poison Ink, which uses colorized image structures as the trigger pattern and hides the trigger pattern invisibly by using a deep injection network.
		\item Extensive experiments demonstrate the stealthiness and robustness of Poison Ink, which is generally applicable to different datasets and network structures.
		\item Poison Ink works well in different attacking scenarios and has strong resistance to many defense techniques. 
	\end{itemize}

	\section{Related Work}
	
	\subsection{Backdoor Attack}
	Backdoor attack is a classic topic in the system security field, and  Gu \etal \cite{gu2019badnets} first introduced this issue into deep models.
	Based on it, Chen  \etal \cite{chen2017targeted}  proposed a blending injection strategy to attack face recognition systems with less poisoned data. 
	Then, Liu \etal \cite{liu2017trojaning} utilized reverse engineering to generate data and built a strong relationship between the optimized patch trigger with the selected neurons via fine-tuning.  
	Unlike these attack methods based on static and visible trigger patterns, the trigger pattern of Poison Ink is dynamic and invisible.
	
	Besides the aforementioned suspicious attack methods, there are a few attempts at more stealthy backdoor attacks.
	Zhong \etal \cite{zhong2020backdoor} proposed an invisible trigger called  static  perturbation mask (SPM), for instance, a checkboard-like pattern. Similarly, Barni \etal \cite{barni2019new}  replaced the repeated mask with the sinusoidal signal (SIG).  Very recently, Liu \etal \cite{liu2020reflection} presented `` Refool ", which is spurred by the natural phenomenon -- reflectance. However, all trigger patterns aforesaid are unnatural due to their input-agnostic attribute.
	Conversely, our proposed method focuses on an input-aware trigger pattern, which is much harder to be detected. 
	Unlike the attacks above, Li \etal \cite{li2019invisible} proposed generating an invisible attack via a steganography algorithm called least significant bit (LSB) substitution. However, it totally fails on low-resolution datasets like CIFAR-10, and Poison Ink significantly outperforms LSB in the context of robustness to data transformation. We also notice some interesting works \cite{chen2021finefool,na2021adversarial,chen2020rca} explored the incorporation of image structures for evasion attacks and defenses. Different from them, we are the first to leverage image structures for backdoor attacks. 
	\vspace{1em}

	\subsection{Defense against Backdoor Attack}
	To resist backdoor attack, many defense methods have been proposed, which can be roughly categorized into three groups: data-based defense, model-based defense, and meta classifiers.  
	
	For data-based defense, Tran \etal \cite{tran2018spectral} removed the poisoned examples by analyzing the spectrum of latent features. However, its assumption of having full access to the infected training data is not practical in use and thus not considered in our experiment.
	Gao \etal's observation included predicting the backdoor image under strong perturbations, on which \textit{STRIP} \cite{gao2019strip} was proposed.
	In addition, Doan \etal \cite{doan2019februus} proposed \textit{Februus} by using Grad-CAM \cite{selvaraju2017grad} to locate the potential trigger region and replacing it by image restoration. Recently, Li \etal \cite{li2020rethinking} showed that most existing attack methods are vulnerable to data transformations, which we mainly focus on in this paper.  
		
	For model-based defense, \textit{Fine-pruning} \cite{liu2018fine} tries to prune the neurons that are dormant for clean inputs, which is assumed to have a relationship to the activation of the backdoor.
	\zj{Chen \etal proposed to detect backdoor attacks by \textit{Activation Clustering} \cite{chen2018detecting}, which determines the infected category by analyzing the activation clustering of all classes.}
	Inspired by the Electrical Brain Stimulation technique, Liu \etal proposed \textit{ABS} \cite{liu2019abs} to scan malicious neuron.
	Then, Wang \etal \cite{wang2019neural} proposed \textit{Neural Cleanse}, which first reverse-engineers the trigger pattern and then utilizes the reversed trigger for backdoor removal. 
	Based on  \textit{Neural Cleanse}, \textit{TABOR} \cite{guo2019tabor} obtained a further improvement by appending various regularization during reverse-engineering. Rather than using reversed trigger, \textit{MESA} \cite{qiao2019defending} leveraged many generated triggers to improve the backdoor removal performance. Later, a more effective method named \textit{TND} is designed by Wang \cite{wang2020practical}, which is applicable even in data-limited or data-free scenarios. 
	
	There are also some methods, such as  \textit{ULPs} \cite{kolouri2020universal}, based on the idea of a meta classifier. However, it consumes enormous computation resources and has a strong but impractical assumption that the trigger size is known. To evaluate the robustness of Poison Ink, we will try these state-of-the-art defense methods in the following experiments.
	
	
	\begin{figure*}[t]
		\hspace{1em}
		\includegraphics[width=0.95\linewidth]{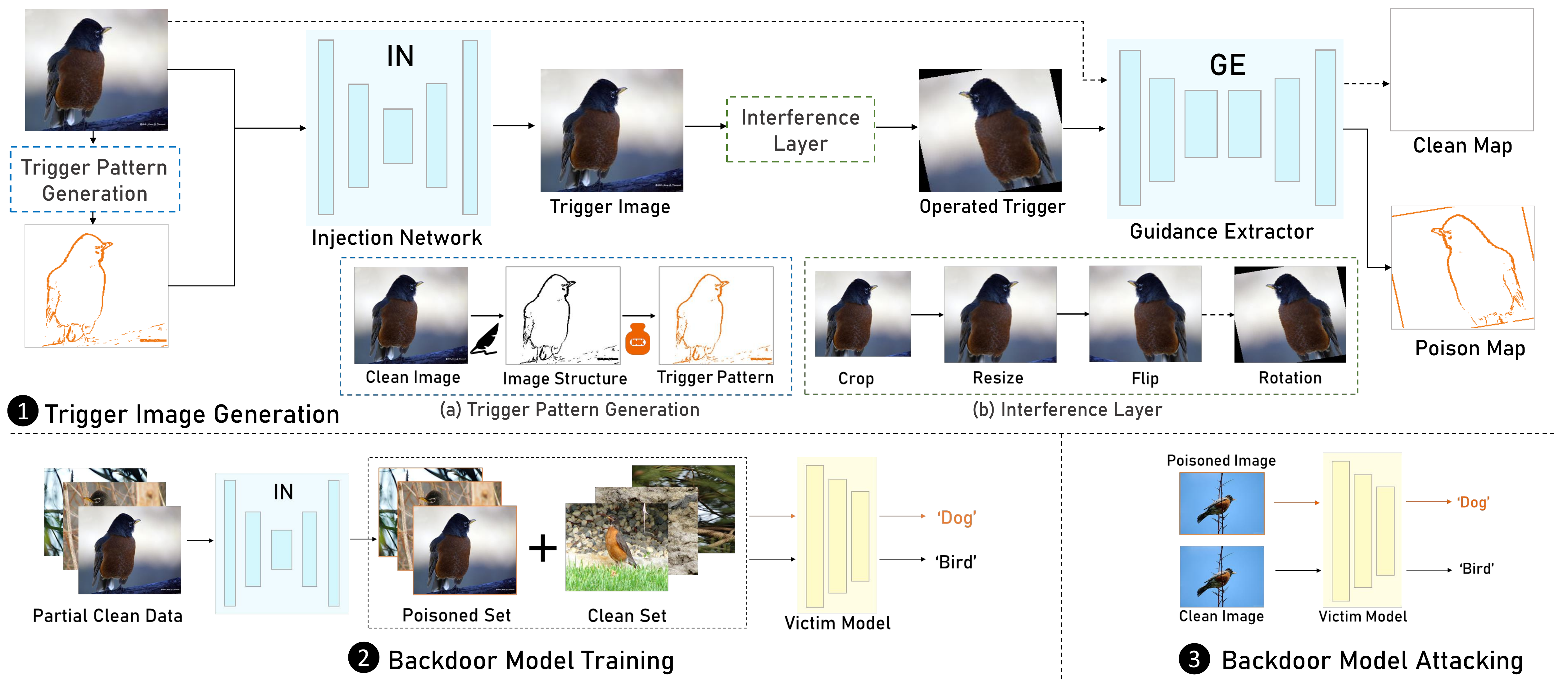}
		\caption{The overall pipeline of Poison Ink, which mainly consists of trigger image generation, backdoor model training and backdoor model attacking. 
		}
		\label{fig:fm}
	\end{figure*}

	\section{Preliminaries}
	Before introducing Poison Ink, we first formally define backdoor attacks and briefly analyze existing methods' limitations. Finally, we clarify the threat model and our attack goals. 
	
	\subsection{Problem Definition}
	In this paper, we only consider the backdoor attack on the image classification task. For image classification, assuming the input domain $\mathbb{X}$ is composed of massive images $\{x_1, x_2, ..., x_N\}$, and the target output domain  $\mathbb{L}$ consists of corresponding labels $\{l_1, l_2, ..., l_N\}$. Then the goal of the image classification model $\mathbf{M}$ is to approximate the implicit transformation function by minimizing the distance $\mathcal{D}$ (eg., cross-entropy) between $\mathbf{M}(x_i)$ and $l_i$, i.e.,
	\begin{equation}
	\mathcal{D}(\mathbf{M}(x_i), l_i) \rightarrow 0.
	\end{equation}
	
	For backdoor attacks, we randomly choose a portion of training data as the candidate set $\mathbb{X}^{c}$ and select some trigger patterns from the pre-designed trigger pattern set $\mathbb{P}$. Here, we pick a single trigger pattern $\mathbf{p}$ from $\mathbb{P}$ as an example.  With a pre-defined backdoor injection strategy $\mathcal{I}$, we can generate the poisoned image $x^{p}$, i.e.,
	\begin{equation}
	x_i^{p} = \mathcal{I} (x_i^{c}, \mathbf{p}), x_i^{c} \in \mathbb{X}^{c}.
	\end{equation}

	All poisoned images $\{x^p_1, x^p_2, ..., x^p_{N_p} \}_{p\in\mathbb{P}}$ with the corresponding target attack labels $\{l_1^p,...,l^p_{N_p}\}$ will be combined with the left clean images $\{x_1, x_2, ..., x_{N_c} \}$ and their original labels $\{l_1, l_2, ..., l_{N_c}\}$  as the final backdoor training dataset $\mathbb{X}^{*}$, where $(\sum_{p\in\mathbb{P}}N_p) + N_c = N $. The injection ratio $\alpha$ is defined as $\alpha = (\sum_{p\in\mathbb{P}}N_p)/N$.
	Finally, we can obtain an infected model $\mathbf{M^*}$ by training on the polluted dataset $\mathbb{X}^{*}$.

	\subsection{Brief Analysis of Existing Limitations}
	Most existing backdoor attack methods have a limitation in stealthiness or robustness. 
	For the stealthiness limitation, we attribute it to two points: the unnatural and input-agnostic trigger patterns; the injection strategy $\mathcal{I}$, like hard-pasting or soft-blending, which is hard to adapt to various inputs to satisfy the invisibility requirement.   
	For the lack of robustness to data transformation, it is also because the trigger pattern $\mathbf{p}$ has no relationship with its cover input $x^{c}$. During training time, the infected model $\mathbf{M^*}$ is forced to remember the relationship between such input-agnostic pattern $\mathbf{p}$ and the target attack label $l^p$. Then, at the inference stage, if the poisoned image $x^{p}$ is pre-processed by some data transformations, the hidden trigger pattern $\mathbf{p}$ will also be transformed, and the relationship will be corrupted.
	Therefore, the infected model $\mathbf{M^*}$ cannot recognize $\mathbf{p}$ any more.

	\subsection{Threat Model and Our Goals}
	Backdoor attacks can occur at any stage of a deep learning pipeline. In this paper, we introduce the threat model in terms of attacker's capacities and attack scenarios as follows:
	
	\noindent\textbf{Attacker's capacities:} we assume that attackers are allowed to poison some training data, whereas they have no information on or change other training components (e.g., training loss, training schedule, and model structure). In the inference process, attackers can and only can query the trained model with any image. They have neither information about the model (even prediction) nor can they manipulate the inference process, \textit{which is different from adversarial example}. 
	The assumption above is the minimal requirement for attackers \cite{li2020backdoor}. Taking data transformations as an example, the attacker is not accessible to the exact range and combinations order of the data transformations used by the defender.
	
	\noindent\textbf{Attack scenarios:} the discussed threat can happen in many real-world scenarios, including but not limited to adopting third-party training data, training platforms, and model APIs. The attacker can directly inject his own poisoned data into the training stage. During inference, the attacker \textit{does not need to hijack any image} and triggers the attack by querying the target model, with the poisoned image generated by himself in the same way.	
	We shall point out that we mainly focus on the threat in the digital domain, where many backdoor attacks were explored, and attacks in the physical world will be our future direction.
	
	We aim to achieve an invisible, robust, general and flexible backdoor attack (Poison Ink) and set main goals in detail: 1) maintain the model performance on clean data; 2) the poisoned image shall be imperceptible to evade human inspection at the inference stage; 3) Keep the high attack effectiveness even when some data transformations pre-process the poisoned image.


	\section{The Proposed Poison Ink} 
	To achieve our goal, we propose a new backdoor attack method called ``Poison Ink'', and the overall pipeline is shown in \Fref{fig:fm}. We are motivated by the recent work \cite{zhang2021exploring} that proposes the physical consistency for model IP protection. When generating the trigger set, we first generate the trigger pattern by embedding the poison information into edge structures and then embed the trigger pattern into the cover image with a deep invisible injection strategy. 
	An interference layer is added to enhance the robustness further and dynamically generates diverse interference to the injection network and the auxiliary guidance extractor network.

	\subsection{Trigger Pattern Generation}
	As described before, the edge structure of one image is an ideal carrier for the trigger pattern: 1) It can be easily captured by the shallow layers of deep models and will not undermine the performance of the original task. 2) The edge structure can keep its semantic meaning and physical existence during data transformation. 3) Different from existing pattern designs, edge structure is also the inherent high-frequency component of an image, so it is easy to be hidden in an invisible way.
	
	In the trigger pattern generation module of \Fref{fig:fm}, we give a simple example to illustrate how to generate a trigger pattern by using the edge structure. In details, given an input image $x_i$, we first outline its edge  by using some edge extraction algorithms $\mathcal{E}$ such as  Sobel \cite{duda1973pattern} or Canny \cite{canny1986computational} operator. Next, we encode the poison information into the RGB color value $\mathcal{C}^p_i$ by mathematical encoding and colorize the black and white edge image with $\mathcal{C}^P_j$ as the target trigger pattern $p_i$:
	\begin{equation}
	p_i = \mathcal{E}(x_i) \otimes \mathcal{C}^p_i,
	\end{equation}
	where $\otimes$ means the color filling operator that makes all the edge color values be $\mathcal{C}^p_i$. Compared to existing trigger patterns, such edge structure based trigger pattern is input-aware and dynamic, which breaks the assumptions of many existing defense techniques that the trigger patterns are input-agnostic and static patches. 
	Besides, since the RGB color space is enormous and many different edge extraction algorithms exist, it also naturally supports multiple different trigger patterns by changing the color values or edge types.

	\subsection{Deep Invisible Injection Strategy}
	After getting the edge structure based trigger pattern, we design a deep injection strategy to hide the trigger pattern into the cover image. As shown in the top part of \Fref{fig:fm}, it basically consists of three parts during training: a deep injection network, an auxiliary guidance extractor network to help the injection network learn in a proper way, and an interference layer to force the injection network to embed trigger pattern more robustly. 
	After the training process, both the auxiliary guidance network and the interference layer will be discarded. We will only use the deep injection network to hide the trigger pattern into the clean cover images to generate the poisoned images.
	
	\vspace{1mm}
	\noindent\textbf{Injection network $\mathbf{\mathit{IN}}$.}  
	We concatenate the clean image $x^c_{i}$ with its corresponding trigger pattern $\mathbf{p}_{i}$ along the channel dimension before feeding them into $\mathbf{\mathit{IN}}$ to obtain the final poisoned image $x^p_{i}$, i.e,
	\begin{equation}
	x^p_{i}  = \mathbf{\mathit{IN}}([x^c_{i};\mathbf{p}_{i}]).
	\end{equation}
	In order to encourage $\mathbf{\mathit{IN}}$ to hide the trigger pattern invisibly, we need to find some invisibility loss metrics to guide its learning. However, it is often difficult to explicitly define invisibility. As shown in existing information hiding methods \cite{zhu2018hidden,zhang2020model,fang2020deep,zhang2021deep}, though $L_p$ Norm is not a perfect invisibility metric, it can serves as a good invisibility learning metric. So we utilize it as one invisibility loss:
	\begin{equation}
	\mathcal{L}_{inv} = \underset{x^c_{i} \in \mathbb{X}^{c}} {\mathbb{E}} \lbrack \lVert \ x^p_{i} - x^c_{i} \rVert^k\rbrack.
	\end{equation}
	\zj{By default, we use L1 loss by setting $k=1$, which follows the classic image-to-image translation framework Pix2Pix \cite{isola2017image}.} To further improve the invisibility, we leverage an extra adversarial loss $\ell_{adv}$ to minimize the domain gap between $x^p_{i}$ and $x^c_{i}$, formally:
	\begin{equation}
	\begin{aligned}
	\ell_{adv} &= \underset{x^c_{i} \in \mathbb{X}^{c}}{\mathbb{E}} log(\mathit{D}(x^c_{i})) + \underset{x^p_{i} \in \mathbb{X}^{p}}{\mathbb{E}} log(1 - \mathit{D}(x^p_{i})). \\
	\end{aligned}
	\end{equation}
	With the adversarial loss, the adversarial discriminator network $\mathit{D}$ will act as a competitor to find the difference between $x^p_{i}$ and $x^c_{i}$. Meanwhile, the injection network $\mathit{IN}$ tries to generate the $x^p_{i}$ in a more invisible way so that the discriminator cannot distinguish it from the clean image $x^c_{i}$.

	\vspace{1mm}
	\noindent\textbf{Guidance extractor $\mathbf{\mathit{GE}}$.} Only constrained by the invisibility loss, the injection network $\mathbf{\mathit{IN}}$ will easily ignore $\mathbf{p}_i$ and learn a trivial solution that outputs the original clean input directly. To ensure $\mathbf{p}_i$ to be hidden in $x^c_{i}$, we append an auxiliary guidance extractor network $\mathbf{\mathit{GE}}$ after the injection network to guide the injection process. 
	
	On the one hand, $\mathbf{\mathit{GE}}$ should be able to extract the trigger pattern $p_i$ out if feeding in the poisoned image $x^p_{i}$, which can be regraded as the reverse operation of injection. On the other hand, $\mathbf{\mathit{GE}}$ should not extract any trigger pattern out from the trigger-free images. With these two requirements, the training of $\mathbf{\mathit{IN}}$ and $\mathbf{\mathit{GE}}$ is indeed conducted in an adversarial way, so it is impossible for $\mathbf{\mathit{IN}}$ to degrade into a trivial solution. To achieve these two goals, we add two corresponding loss functions respectively: the trigger extraction loss $\mathcal{L}_{te}$ for poisoned images and the clean loss $\mathcal{L}_{cl}$ for trigger-free images i.e.,
	
	\begin{equation}
	\begin{aligned}
	\mathcal{L}_{GE} &= \mathcal{L}_{te} +  \lambda \cdot \mathcal{L}_{cl}, \\
	\mathcal{L}_{te} &= \underset{x^p_{i} \in \mathbb{X}^p} {\mathbb{E}} \lbrack \lVert \mathbf{\mathit{GE}}(x^p_{i}) - \mathbf{p}_{i} \rVert_2\rbrack, \\
	\mathcal{L}_{cl} &= \underset{x^c_{i} \in \mathbb{X}^c} {\mathbb{E}} \lbrack \lVert \mathbf{\mathit{GE}}(x^c_{i}) - \mathbf{C} \rVert_2\rbrack,
	\end{aligned}
	\end{equation}
	where $\mathbf{C}$ is a clean map (shown in \Fref{fig:fm}) to indicate no trigger pattern and $\lambda$ is one hyper-parameter to balance the two loss terms. Here we simply use the $L2$ reconstruction loss for both $\mathcal{L}_{te}$ and $\mathcal{L}_{cl}$ by default.

	\vspace{1mm}
	\noindent\textbf{Interference layer.} To increase the extracting difficulty of the guidance extractor network and encourage the injection network to hide the trigger pattern in a more robust way, we further add an auxiliary interference layer between them. During training, this layer will randomly augment the poisoned image output from the injection network. By default, it consists of a sequence of common data augmentation operators $\{ \mathcal{T}_{crop} \rightarrow  \mathcal{T}_{resize} \rightarrow \mathcal{T}_{flip} \rightarrow \cdots \rightarrow \mathcal{T}_{rot} \}$, and each poison image will be randomly augmented by each operator with the probability of 0.5. 
	
	To guarantee both the invisibility and robustness, the injection network $\mathbf{\mathit{IN}}$ and the guidance network  $\mathbf{\mathit{GE}}$ are jointly trained, and the total loss function $\mathcal{L}_{total}$ is: 
	\begin{equation}
	\mathcal{L}_{total} = \mathcal{L}_{IN} +  \gamma \cdot \mathcal{L}_{GE}.
	\end{equation}
	By default, we set $\gamma=1$. During training, we will randomly use different color values $\mathcal{C}^p_i$ for different image $x_i$ and encourage $\mathbf{\mathit{IN}}$ to be a general injection network that can embed different types of trigger patterns on the fly. When attacking one target model, we can choose one specific trigger pattern or multiple different trigger patterns by using different $\mathcal{C}^p_i$ to enable single-label attack and multi-label attack, respectively.

	\section{Experiments}
	In this section, we will first briefly introduce the implementation details we adopt, such as datasets, network structures, metrics, and default settings. After that, we will demonstrate the invisibility and robustness of the proposed Poison Ink, respectively. Then more network structures, datasets, and attack scenarios are considered to demonstrate the generality and flexibility of our method. Next, we showcase the resistance of Poison Ink to many state-of-the-art backdoor defense methods. Finally, some ablation studies are conducted to justify our design. Our source code will be released.
	
	\subsection{Implementation Details.}
	\noindent\textbf{Datasets.} 
	We consider 4 datasets for three types of classification tasks. For  object recognition, we use CIFAR-10  \cite{krizhevsky2009learning} and ImageNet \cite{deng2009imagenet}; GTSRB \cite{stallkamp2011german} and VGG-Face \cite{parkhi2015deep} are utilized for traffic sign recognition and face recognition, respectively. For each dataset except CIFAR-10, we randomly select a portion of classes and resize the inputs for diversity, as shown in \Tref{tab:dataset}.  
	Because of the resource and space consideration, we mainly use the ImageNet and CIFAR-10 for comparison and CIFAR-10 for ablation. GTSRB and VGG-Face are used to demonstrate the generality of our method. Note that, \zj{The test data has no overlap with the training data, and the trigger pattern is input-aware and not seen during the training phase.}
	
	\vspace{0.4em}
	\noindent\textbf{Network structures.} For trigger image generation, we simply adopt the UNet \cite{ronneberger2015u} and the PatchGAN \cite{pix2pix2017} as the default network structure of $\mathbf{\mathit{IN}}$ and the discriminator $\mathit{D}$ respectively, which are both widely used in many image-to-image tasks\cite{isola2017image,zhu2017unpaired,chen2020controllable}. For guidance extractor $\mathit{GE}$, we design a simple auto-encoder-like network. The encoder consists of three residual blocks with stride 2, symmetrically, and the decoder consists of three residual upsampling blocks to ensure the output resolution to be same as the input resolution. To further enhance its learning capacity, several residual blocks are also inserted between the encoder and the decoder. 
	In \Tref{tab:in} and \Tref{tab:ge}, we provide the details of the injection network $\mathit{IN}$ and the guidance extractor $\mathit{GE}$ respectively. 
	Besides, we chop up each poisoned image into 16 $\times$ 16 patches for the discriminator $\mathit{D}$.

	As for network structures of the victim classifier, we consider four popular recognition networks: ResNet-18 \cite{he2016deep}, ResNeXt \cite{xie2017aggregated}, DenseNet \cite{huang2017densely} and VGG-19 \cite{simonyan2014very}. In the following experiments, we adopt VGG-19 as the default network and mainly showcase the results on CIFAR-10 and ImageNet. The results of other network structures and datasets are used to demonstrate the generality of our method.
	
	\begin{table}[t]
		\centering
		\caption{\zj{The statics of datasets.}}
		\setlength{\tabcolsep}{1mm}{\begin{tabular}{c|c|c|c}
				\hline
				Dataset  & \# Labels & Input Size & \# Training / Testing Images \\ \hline
				CIFAR-10  & 10        & 32 $\times$ 32    & 50000 / 5000              \\ \hline
				GTSRB    & 13        & 64 $\times$ 64    & 4772 / 293              \\  \hline         
				ImageNet & 100       & 224 $\times$ 224  & 126689 / 5000            \\ \hline
				VGG-Face & 500       & 224 $\times$ 224  & 135712  / 3147           \\ \hline
		\end{tabular}}
		\label{tab:dataset}
	\end{table}
	
	\begin{table*}[!h]
		\centering
		\caption{The detailed network structure of the injection network $\mathit{IN}$.}		
		\setlength{\tabcolsep}{2.6mm}{	
			\begin{tabular}{c|cccc|c|c|c}
				\hline
				Module   Name                    & Layer Name       & Kernel & Stride & Channel I/O  & Normalization & Activation & Input                    \\ \hline
				\multirow{5}{*}{Encoder}
				& conv1            & 4 x 4  & 2    & 6/64     & BN            & LeakyReLU       & Cat[(Images, Trigger Pattern), 1]                  \\  
				& conv2            & 4 x 4  & 2    & 64/128   & BN            & LeakyReLU       & conv1                 \\ 
				& conv3            & 4 x 4  & 2    & 128/256  & BN            & LeakyReLU       & conv2                 \\
				& conv4            & 4 x 4  & 2    & 256/512  & BN            & LeakyReLU       & conv3                 \\
				& conv5            & 4 x 4  & 2    & 512/512  & BN            & LeakyReLU       & conv4                 \\ \hline
				\multirow{5}{*}{Decoder}  		  
				& transpose\_conv1            & 4 x 4  & 2    & 512/512      & BN            & ReLU       & conv5                   \\ 
				& transpose\_conv2            & 4 x 4  & 2    & 1024/256     & BN            & ReLU       & transpose\_conv1 + conv4                   \\ 
				& transpose\_conv3            & 4 x 4  & 2    & 512/128      & BN            & ReLU       & transpose\_conv2 + conv3                  \\ 
				& transpose\_conv4            & 4 x 4  & 2    & 256/64       & BN            & ReLU       & transpose\_conv3 + conv2                   \\ 
				& transpose\_conv5            & 4 x 4  & 2    & 128/3        & BN            & Sigmoid    & transpose\_conv4 + conv1                   \\ \hline
		\end{tabular}}
		\label{tab:in}
	\end{table*}

	\begin{table*}[h]
		\centering
		\caption{The detailed network structure of the guidance extractor $\mathit{GE}$.}
		\setlength{\tabcolsep}{3.5mm}{
			\begin{tabular}{c|cccc|c|c|c}
				\hline
				Module   Name                            & Layer Name  & Kernel & Stride & Channel I/O & Normalization & Activation & Input          \\ \hline
				\multirow{3}{*}{Encoder } 
				& conv1       & 3 x 3  & 1    & 3/128   & IN            & ReLU       & Poisoned Images         \\ 
				& conv2       & 3 x 3  & 1    & 128/128   & IN            & ReLU       & conv1         \\ 
				& conv3       & 3 x 3  & 2    & 128/128   & IN            & ReLU       & conv2         \\ \hline
				\multirow{7}{*}{Residual Learning}     
				& res\_block1 & 3 x 3  & 1    & 128/128   & IN            & ReLU       & conv3       \\ 
				& res\_block2 & 3 x 3  & 1    & 128/128   & IN            & ReLU       & res\_block1 \\  
				& res\_block3 & 3 x 3  & 1    & 128/128   & IN            & ReLU       & res\_block2 \\
				& res\_block4 & 3 x 3  & 1    & 128/128   & IN            & ReLU       & res\_block3 \\ 
				& res\_block5 & 3 x 3  & 1    & 128/128   & IN            & ReLU       & res\_block4 \\ 
				& res\_block6 & 3 x 3  & 1    & 128/128   & IN            & ReLU       & res\_block5 \\  
				& res\_block7 & 3 x 3  & 1    & 128/128   & IN            & ReLU       & res\_block6 \\ \hline
				\multirow{3}{*}{Decoder} 
				& transpose\_conv1       & 4 x 4  & 2    & 128/128   & IN             & ReLU       & res\_block7 \\ 
				& transpose\_conv2       & 3 x 3  & 1    & 128/128   & IN             & ReLU       & transpose\_conv1  \\ 
				& transpose\_conv3       & 1 x 1  & 1    & 128/3     & IN             & -      & transpose\_conv2  \\ \hline
		\end{tabular}}
		\label{tab:ge}
	\end{table*}

	\vspace{0.2em}
	\noindent\textbf{Metrics.} We use Clean Data Accuracy (CDA) to evaluate the influence of backdoor attacks on the original tasks, and use Attack Success Rate (ASR) to evaluate the effectiveness of backdoor attacks. Specifically, CDA is the performance on the clean test set, \ie, the ratio of trigger-free test images that are correctly predicted to their ground-truth labels; and ASR is the performance on the pre-defined poisoned test set, \ie, the ratio of poisoned images that are correctly classified as the target attack labels. 
	
	For invisibility evaluation, we compare clean and poisoned images with three famous metrics, PSNR, SSIM, and LPIPS, where LPIPS adopts the features of the pre-trained AlexNet. Besides the above metrics, we also conduct a user study for human inspection testing.

	\vspace{0.2em}
	\noindent\textbf{Default settings for training.} For trigger pattern generation, we use the Sobel operator \cite{duda1973pattern} to extract the edge. We conduct comparisons with the single-label attack by default and inject the poison ink (R:80, G:160, B:80) into the edge area with the well-trained injection network  $\mathbf{\mathit{IN}}$.  Then, 10\% pollution rate is considered for all tasks, and the first class of each dataset is chosen as the target attack label. \zj{We train the injection network  $\mathbf{\mathit{IN}}$ for 200 epochs by default.}
	All the victim models are trained using the SGD optimizer with a momentum of 0.9 and an initial learning rate of 0.01, which is further set as 0.001 and 0.0001 at epoch 150 and epoch 200, respectively.
	For the other methods to compare with, we adopt the default setting in their official implementations.

	\begin{figure*}[t]
		\hspace{3em}
		\includegraphics[width=0.85\linewidth]{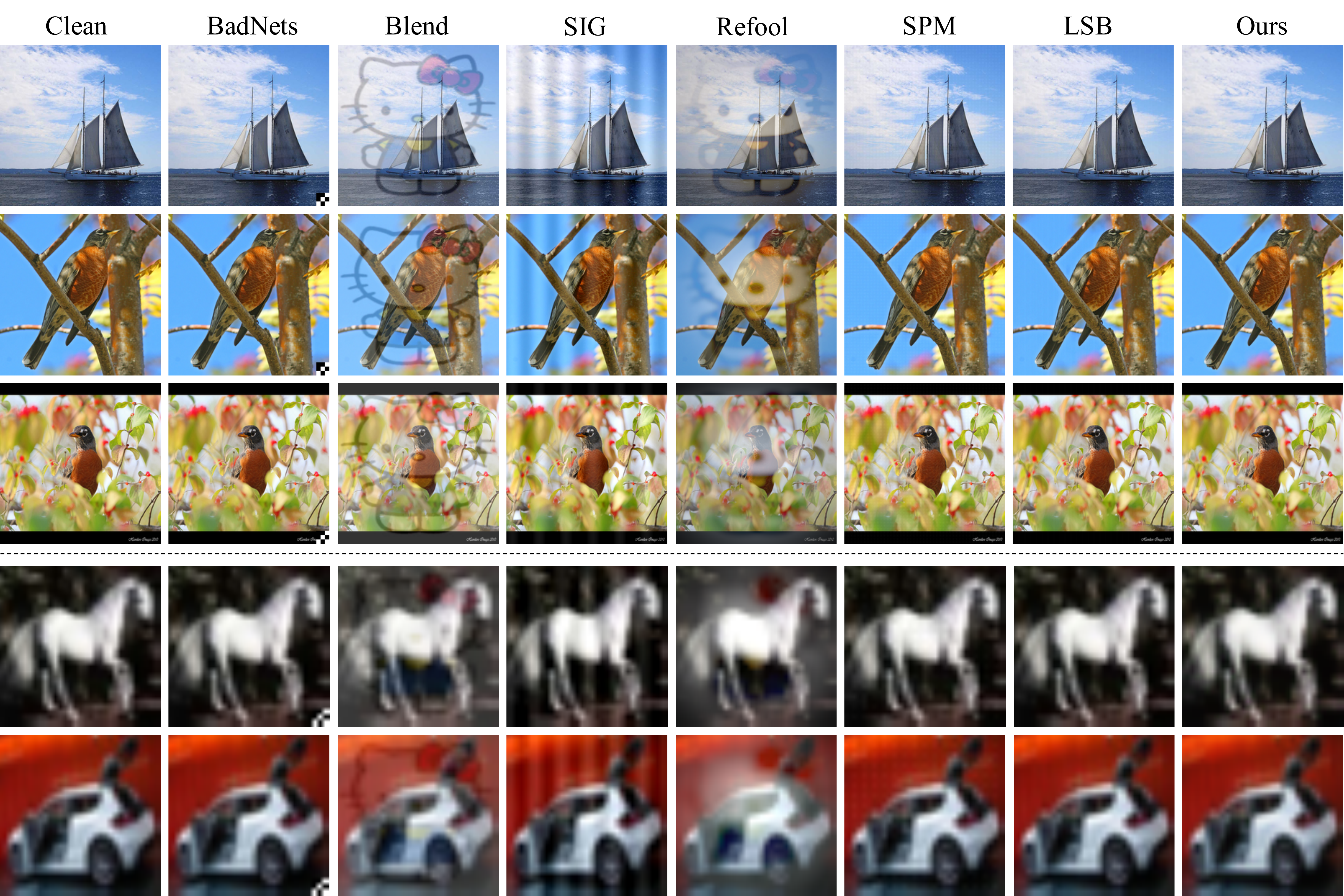}
		\caption{Visual comparison with existing popular attack methods. The first three rows are examples on ImageNet dataset and the last two rows are examples on CIFAR-10 dataset. ``Clean'' denotes the original trigger-free image.
		}
		\label{fig:vis_cmp}
	\end{figure*}  
	
	\begin{table*}[t]
		\centering
		\caption{Comparison of invisibility (stealthiness) with existing popular backdoor attack methods on ImageNet dataset.}
		\label{tab:vis_cmp1}
		\setlength{\tabcolsep}{4.5mm}{ 
			\begin{tabular}{c|c|c|c|c|c|c|c}
				\hline
				Metric   & BadNets \cite{gu2019badnets}& Blend \cite{chen2017targeted} & SIG \cite{barni2019new} &  Refool \cite{liu2020reflection} & SPM \cite{zhong2020backdoor}  &LSB \cite{li2019invisible} &   Ours \\ \hline
				PSNR $\uparrow$ & 27.03      &  19.18     &  25.12        & 16.59  &    38.65        & 51.14   &      41.62     \\ \hline
				SSIM $\uparrow$ & 0.9921     &  0.7921     &   0.8988     & 0.7701  &     0.9665     & 0.9975  &     0.9915  \\ \hline
				LPIPS $\downarrow$ & 0.0149  &  0.2097     &   0.0532   & 0.2461   &     0.0022      & 0.0003  &    0.0020 \\  \hline
				Fooling Rate (\%)  & 0        &  0     & 0      & 0  & 38.33        &   51.67      & 53.33  \\  \hline
		\end{tabular}}
	\end{table*}
	
	\begin{table*}[t]
		\centering
		\caption{Comparison of invisibility (stealthiness) with existing popular backdoor attack methods on CIFAR-10 dataset.}
		\label{tab:vis_cmp2}
		\setlength{\tabcolsep}{4.5mm}{ 
			\begin{tabular}{c|c|c|c|c|c|c|c}
				\hline
				Metric   & BadNets \cite{gu2019badnets}& Blend \cite{chen2017targeted} & SIG \cite{barni2019new} &  Refool \cite{liu2020reflection} & SPM \cite{zhong2020backdoor}  &LSB \cite{li2019invisible} &   Ours \\ \hline
				PSNR $\uparrow$ & 25.68        &  21.29     & 25.12     & 19.38  & 38.94         & 51.13        & 42.95     \\ \hline
				SSIM $\uparrow$ & 0.9833       & 0.8810     & 0.8250    &  0.8243 & 0.9884       & 0.9991      & 0.9961    \\  \hline
				LPIPS $\downarrow$  & 0.0009   & 0.0240     & 0.0400    &  0.0397    & 0.0001     & 0.00001     & 0.0001      \\ \hline
				Fooling Rate (\%)  & 0        &  0     & 0      & 0  & 16.67        & 43.33        & 46.67 				  \\  \hline
		\end{tabular}}
	\end{table*}

	\subsection{Invisibility of Poison Ink}
	We first compare the invisibility of our method with many popular backdoor attack methods, and we follow their default implementation for fair comparisons.  
	
	In \Fref{fig:vis_cmp}, we showcase more visual comparison with other popular attack methods on ImageNet dataset and CIFAR-10 dataset. In detail, we can see that the poisoned image generated by BadNets \cite{gu2019badnets}, Blend \cite{chen2017targeted}, SIG \cite{baluja2017hiding} and Refool \cite{liu2020reflection} can be easily distinguished from the clean images, due to the image-agnostic trigger pattern and simple embedding strategy. As for SPM \cite{zhong2020backdoor}, the checkboard-like pattern is also easily observed in the smooth region of the image. In contrast, LSB \cite{li2019invisible} and Poison Ink  achieve better invisibility, and the embedded poison is imperceptible. 
	
	In \Tref{tab:vis_cmp1} and \Tref{tab:vis_cmp2}, we provide the quantitative comparison of invisibility on ImageNet dataset and CIFAR-10 dataset. Poison Ink outperforms the majority of attacks under all the evaluation metrics except LSB \cite{li2019invisible}, which are consistent with the visual comparison.
	As for LSB, it also has good invisibility but is very fragile to common data transformations in the attacking stage (shown in \Tref{tab:robust1} and \Tref{tab:robust2}). Moreover, in \Tref{tab:robust2}, we observe that LSB fails on CIFAR-10 dataset, even without transformation-based pre-processing on its poisoned images. 	It may be because the trigger pattern of LSB on the CIFAR-10 dataset is regarded as random noise and ignored by the target classifier.

	We further conduct a user study for human inspection testing. In detail, we randomly selected 50 clean images from ImageNet dataset and CIFAR-10 dataset and generated the corresponding 50 poisoned images for each backdoor attack method. Then in each question, we randomly display one image pair (one clean image and one \zj{corresponding} poisoned image) and ask the user to select which one is the clean image. A total of 30 volunteers \zj{(12 females and 18 males, user ages range from 18 to 30, and all users are familiar with backdoor attacks)} are involved in our user study. Thus, there are 1500 answers for each attack method.  We compare the fooling rate of each method as shown in the last row of \Tref{tab:vis_cmp1} and \Tref{tab:vis_cmp2}. It can be observed that the fooling rate of LSB and our method both are close to 50\%, a probability of random guessing, while the poisoned images generated by other remaining methods can be easily judged as unclean.

	\subsection{Influence on Pristine Performance}
	Besides the stealthiness of poisoned images, another critical aspect for evaluating the backdoor attack is its influence on pristine performance. In other words, when training the backdoor model with the mixture of the trigger set and the clean set, the model should keep its original performance on the clean set. In the left part of \Tref{tab:robust1} and \Tref{tab:robust2}, we show the Clean Data Accuracy (CDA) of backdoored models trained with the trigger set generated by different methods. It can be seen that the backdoored models of Poison Ink have overall higher CDA than other baselines, demonstrating the stealthiness during the backdoor model training. Even in the worst case, the CDA is still comparable with other methods. Here, the CDA of the original model without backdoor attack is 80.84\% on ImageNet dataset and 91.41\% on CIFAR-10 dataset, respectively.

	\begin{table*}[t]
		\centering
		\footnotesize
		\caption{Comparison of the robustness against different data transformations with popular attack methods on  ImageNet dataset. ``DT'' denotes data transformation. ``S\&P'' and ``C\&R'' means padding after shrinking and resizing after cropping, respectively. \zj{The red font represents the "worst-case" performance.} }   		
		\setlength{\tabcolsep}{2.8mm}{ 
			\begin{tabular}{c|c|c|c|c|c|c|c|c|c|c|c|c}
				\hline
				\centering{Metric}  & \multicolumn{6}{c|}{CDA (\%)}                       & \multicolumn{6}{c}{ASR (\%)}                        \\ \hline
				DT      & None  & Flip  & S\&P  & Rot 15 & C\&R  & Average   & None   & Flip  & S\&P  & Rot 15 & C\&R  & Average   \\ \hline
				BadNets \cite{gu2019badnets} & 80.78 & 80.32 & \wc{73.90} & 78.52  & 75.82 & 77.87 & 100.00 & 99.98 & 31.46 & \wc{6.08}   & 55.34 & 58.57 \\ \hline
				Blend \cite{chen2017targeted}  & 80.12 & 80.30 & \wc{74.72} & 77.52  & 76.30 & 77.79 & 99.98  & 99.96 & 96.40 & \wc{69.72}  & 89.46 & 91.10 \\ \hline
				SIG  \cite{barni2019new}  & 80.42 & 80.34 & \wc{\textbf{74.82}} & 77.90  & 76.40 & \textbf{77.98} & 100.00 & 99.98 & \wc{51.84} & 54.20  & 68.94 & 74.99 \\ \hline
				Refool \cite{liu2020reflection}  & 80.36 & 80.10 & \wc{74.60} & 78.24  & 75.84 & 77.83 & 99.36  & 99.32 & 98.70 & \wc{\textbf{92.00}}  & 92.24 & \textbf{96.32} \\ \hline
				SPM  \cite{zhong2020backdoor}   & 80.42 & 80.10 & \wc{73.16} & 77.60  & 76.10 & 77.48 & 99.94  & 99.92 & 53.94 & \wc{1.04}   & 0.90  & 51.15 \\ \hline
				LSB  \cite{li2019invisible}   & 80.64 & 80.28 & \wc{73.60} & 77.82  & 75.98 & 77.66 & 97.32  & 97.22 & 11.12 & \wc{0.92}   & 0.98  & 41.51 \\ \hline
				Ours    & 80.56 & 80.40 & \wc{\textbf{74.32}} & 78.10  & 76.64 & \textbf{78.00} & 98.48  & 98.62 & 98.20 & \wc{\textbf{96.10}}  & 96.12 & \textbf{97.50} \\ \hline
		\end{tabular}}
		\label{tab:robust1}
	\end{table*}
	
	\begin{table*}[t]
		\centering
		\caption{Comparison of the robustness against different data transformations with popular attack methods on  CIFAR-10 dataset. ``DT'' denotes data transformation. ``S\&P'' and ``C\&R'' means padding after shrinking and resizing after cropping, respectively.  \zj{The red font represents the "worst-case" performance.}} 
		\setlength{\tabcolsep}{2.8mm}{ 
			\begin{tabular}{c|c|c|c|c|c|c|c|c|c|c|c|c}
				\hline
				\centering{Metric}  & \multicolumn{6}{c|}{CDA (\%)}                       & \multicolumn{6}{c}{ASR (\%)}                        \\ \hline
				DT      & None  & Flip  & S\&P  & Rot 15 & C\&R  & Average   & None   & Flip  & S\&P  & Rot 15 & C\&R  & Average   \\ \hline
				BadNets \cite{gu2019badnets}  & 91.22 & 91.15 & 86.01 & \wc{77.76} & 81.08 & 85.44 & 100   & 99.98 & 22.00    & 77.26 & \wc{39.77} & 67.80 \\ \hline
				Blend \cite{chen2017targeted}   &93.32 & 93.01 & 89.89 & 86.97 & \wc{85.28} & \textbf{89.69} & 99.82 & 99.79 & 92.39 & 98.95 & \wc{74.34} & 93.06 \\ \hline
				SIG  \cite{barni2019new}   & 92.64 & 93.08 & 89.34 & 86.23 & \wc{\textbf{85.43}} & 89.34 & 99.92 & 99.95 & 99.86 & 99.8  & \wc{\textbf{97.62}} & \textbf{99.43}  \\ \hline
				Refool  \cite{liu2020reflection} & 92.40 & 92.57 & 88.72 & 86.20 & \wc{83.34} & 88.65 & 93.87 & 93.64 & \wc{89.59} & 93.22 & 93.63 & 92.79  \\ \hline
				SPM \cite{zhong2020backdoor}    &93.05 & 93.39 & 89.63 & \wc{84.51} & 84.75 & 89.07 & 99.86 & 99.89 & 9.72  & 68.42 & \wc{8.96}  & 57.37  \\ \hline
				LSB \cite{li2019invisible}     & 87.86 & 87.22 & 84.28 & 81.26 & \wc{77.54} & 83.63 & \wc{14.93} & 15.36 & 15.14 & 16.35 & 17.03 & 15.76 \\ \hline
				Ours    & 92.92 & 92.77 & 89.60 & 87.09 & \wc{\textbf{85.15}} & \textbf{89.51} & 99.92 & 99.97 & \wc{\textbf{83.12}} & 96.94 & 87.46 & \textbf{93.48} \\ \hline
				
		\end{tabular}}
		\label{tab:robust2}
	\end{table*}

	\begin{figure*}[!t]
		\hspace{3em}
		\includegraphics[width=0.9\linewidth]{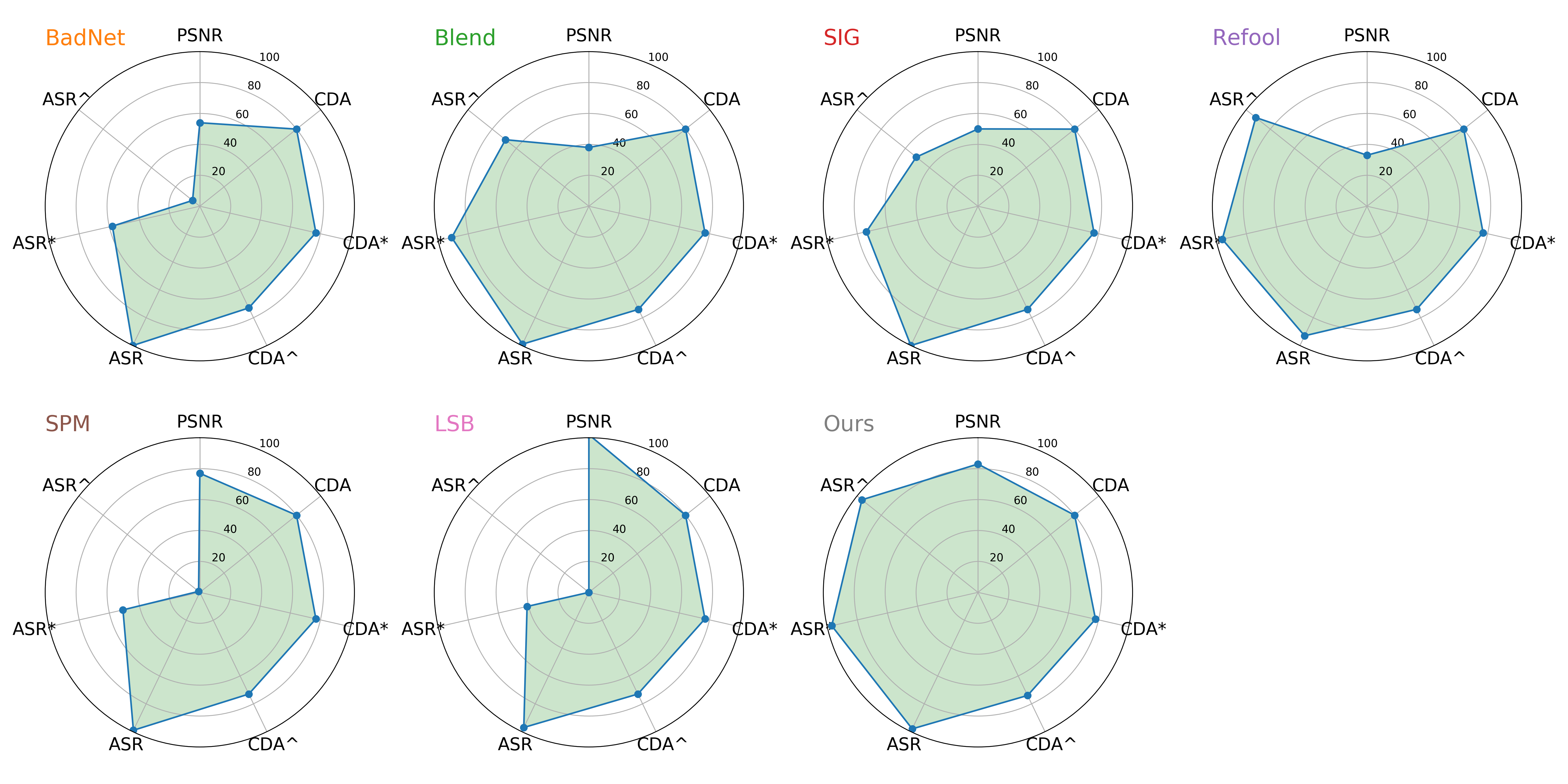}
		\caption{Comparison with popular attack methods on ImageNet dataset in terms of both stealthiness and robustness. PSNR is 2$\times$ original value. CDA and ASR mean the performance under no data transformation, CDA$^*$ and ASR$^*$ denote the average performance under different data transformations, \zj{and CDA$^{\wedge}$ and ASR$^{\wedge}$ are calculated under the worst case.}}
		\label{fig:ra_in1}
	\end{figure*} 
	
	\begin{figure*}[!t]
		\hspace{3em}
		\includegraphics[width=0.9\linewidth]{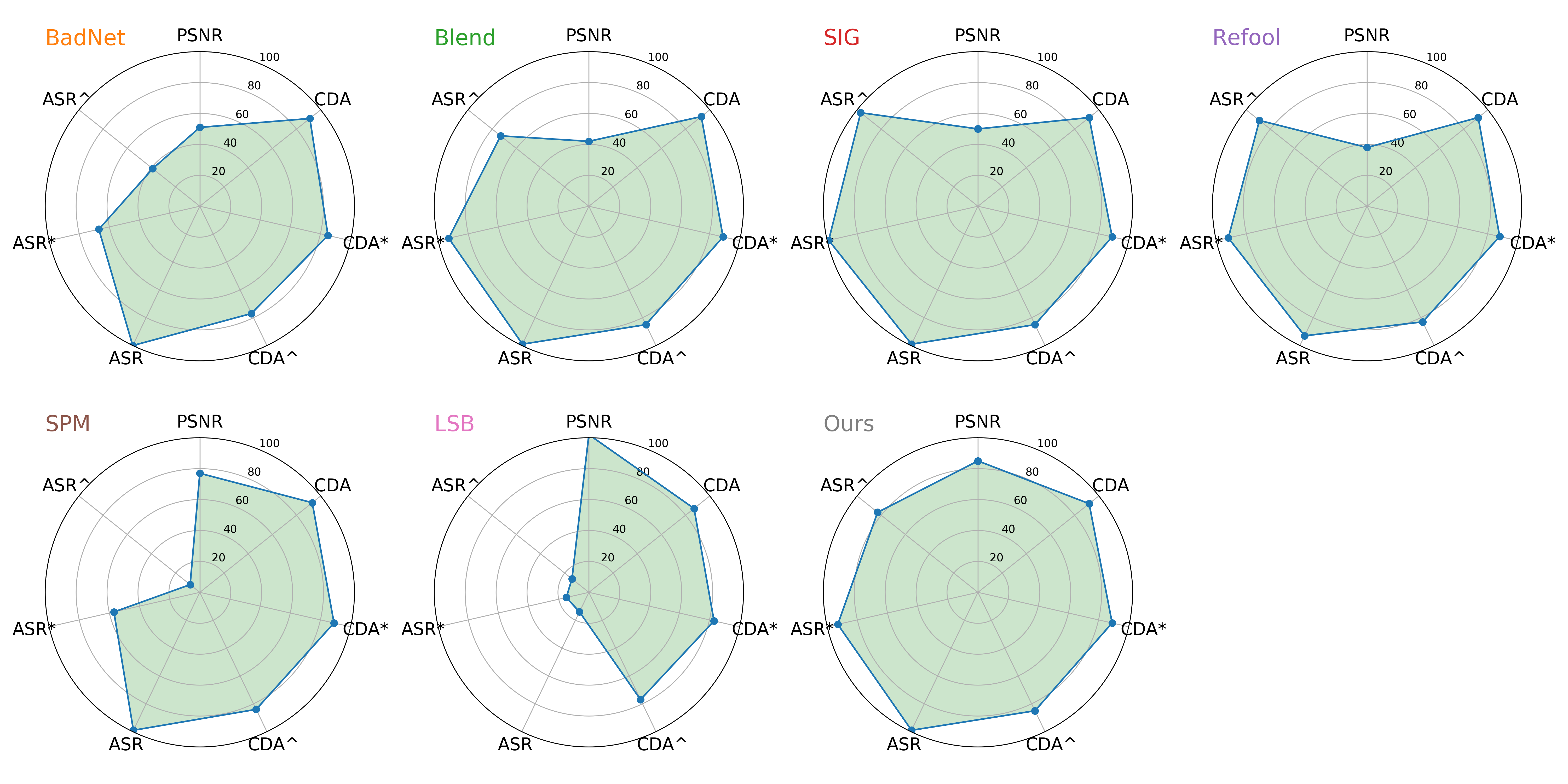}
		\caption{Comparison with popular attack methods on CIFAR-10 dataset in terms of both stealthiness and robustness. PSNR is 2$\times$ original value.CDA and ASR mean the performance under no data transformation, CDA$^*$ and ASR$^*$ denote the average performance under different data transformations, \zj{and CDA$^{\wedge}$ and ASR$^{\wedge}$ are calculated under the worst case.}}
		\label{fig:ra_in2}
	\end{figure*}

	\subsection{Robustness of Poison Ink}
	To address the limitation to data transformation, a corresponding solution \cite{li2020rethinking} was proposed that conducting a data transformation on the training images with the trigger before feeding into the training process, which can be seen as adversarial training. In our threat model, the attacker cannot control the training strategy. Nevertheless, in this section, we train all backdoor attacks in such an enhanced training strategy for a fair comparison.  
	
	In the right part of \Tref{tab:robust1}, we give the ASR of different attack methods on ImageNet dataset without (None) or with data transformation attack, respectively. With the data transformation attack, the poisoned images will be pre-processed by different data transformation techniques and then fed into the victim model for testing. \zj{We also provide the results in the average case and the worst case in order to evaluate these attack methods comprehensively}.
	It can be seen that, compared with other popular attack methods, Poison Ink achieves a higher ASR in both the average and worst cases, when facing different transformations. In contrast, some baseline methods like SPM and LSB will totally fail when some specific transformations are applied. Compared to Refool, our method outperforms narrowly in terms of robustness but outperforms in stealthiness by a large margin, as shown in \Tref{tab:vis_cmp1}. 

	Although the enhanced training strategy improves the robustness to some extent, overfitting will be introduced. Taking SPM as an example, such a strategy improves the robustness against rotation but fails on other data transformations. We explain that SPM tends to only focus on getting robust against rotation. Conversely, Poison Ink mitigates such overfitting and achieves an outstanding or comparable performance in terms of the CDA and the ASR.
	
	As shown in \Tref{tab:robust2}, we provide the quantitative comparison of robustness to data transformations on CIFAR-10 dataset. Overall, Poison Ink guarantees comparable robustness on CIFAR-10 dataset with many existing attack methods such as Blend and SIG, which nevertheless have worse invisibility compared with our method (shown in \Fref{fig:vis_cmp} and \Tref{tab:vis_cmp2}). In \Fref{fig:ra_in1} and \Fref{fig:ra_in2}, we plot radar charts to illustrate that our method achieves a much better balance between visual quality and robustness.

	\subsection{Comparison with More Invisible Backdoor Attacks}
	\zj{Besides the classic backdoor attacks mentioned above, we further consider some recent invisible backdoor attacks, such as WaNet \cite{nguyen2021wanet}, FTrojan \cite{wang2021backdoor}, and Advdoor \cite{zhang2021advdoor}.
	In detail, WaNet generates backdoor images via subtle image warping, and FTrojan injects mid- and high-frequency triggers in each block with medium magnitude, both of which utilize input-aware trigger patterns. Advdoor generates triggers by the targeted universal adversarial perturbation (TUAP). Specifically, some inputs in a category are used to obtain input-specific adversarial perturbations firstly, which are further integrated to generate the final TUAP for the target category.  Such TUAPs belong to the static trigger patterns, which are further directly imposed onto the clean images to generate final triggers.}
	
\begin{figure}[!t]
		\centering
		\includegraphics[width=1\linewidth]{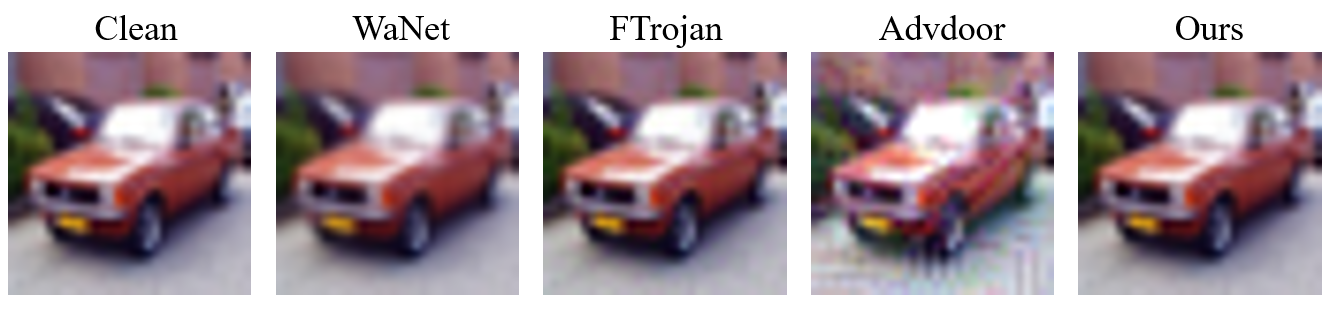} 
		\caption{ \zj{Visual examples of a poisoned image generated by invisible backdoor attack methods.  }}
		\label{fig:invs}
	\end{figure}

	\zj{All comparison experiments are conducted on CIFAR-10 dataset, and we train all backdoored models with the enhanced training strategy \cite{li2020rethinking} for a fair comparison. We calculate the average CDA and ASR under different data transformations. As shown in \Tref{tab:invis} and \Fref{fig:invs}, Advdoor is not stealthy enough due to its static trigger patterns, while WaNet is fragile to data transformations. Compared to Advdoor and WaNet, FTrojan and our method perform well in both stealthiness and robustness. If training backdoored model without the enhanced strategy, poison ink achieves stronger robustness than FTrojan (ASR: 83.67\% vs. 59.51\% in \Tref{tab:ft}). }

	\begin{table}[]
		\centering
		\caption{\zj{The comparison results with recent invisible backdoor attack methods.}	}	
		\setlength{\tabcolsep}{1.8mm}{
			\begin{tabular}{c|c|c|c|c|c}
				\hline
				Methods & CDA(\%)   & ASR(\%)   & PSNR $\uparrow$   & SSIM $\uparrow$    & LIPIS $\downarrow$   \\ \hline
				WaNet \cite{nguyen2021wanet}  & 83.47 & 26.10  & 32.15 & 0.9822 & 0.0042 \\ \hline
				FTrojan  \cite{wang2021backdoor} & 87.47 & 98.21 & 41.10  & 0.9915 & 0.0001 \\ \hline
				Advdoor \cite{zhang2021advdoor}  & 87.69 & 99.71   &22.97  &0.9101  &0.0109 \\ \hline
				ours    & 89.51 & 93.48 & 42.95 & 0.9961 & 0.0001 \\ \hline
		\end{tabular}}
		\label{tab:invis}
	\end{table}

	\begin{table}[]
		\centering
		\caption{ \zjj{Comparison of the robustness against different data transformations with FTrojan  \cite{wang2021backdoor} on  CIFAR-10 dataset. $\dagger$ denotes without the enhanced training strategy.}	}
		\setlength{\tabcolsep}{2mm}{
		\begin{tabular}{c|c|c|c|c|c|c}			
			\hline
			Methods & None  & Flip  & S\&P  & Rot 15 & C\&R  & Average \\ \hline
			FTrojan  \cite{wang2021backdoor}      & 97.57 & 97.35 & 99.07 & 97.59  & 99.46 & 98.21 \\ \hline
			ours     & 99.92 & 99.97 & 83.12 & 96.94  & 87.46 & 93.48 \\ \hline
			FTrojan  \cite{wang2021backdoor} $\dagger$     & 100   & 100   & 9.99  & 77.99  & 9.56  & 59.51 \\ \hline
			ours  $\dagger$   & 99.92 & 99.98 & 39.82 & 96.24  & 82.39 & 83.67 \\ \hline
		\end{tabular}}
	\label{tab:ft}
	\end{table}

	\begin{table*}[!t]
		\caption{The performance (OMA / CDA / ASR  (\%)) of Poison Ink on different datasets and different network structures.
			We adopt VGG-19 and CIFAR-10 as default network architecture and default dataset, respectively. 
			``OMA'' denotes accuracy of the original model trained on total clean dataset.}
		\centering
		\setlength{\tabcolsep}{3.2mm}{ 
			\begin{tabular}{c|cc |ccc}
				\hline
				Settings & GTSRB & VGG-Face  & ResNet-18 & ResNeXt & DenseNet    \\ \hline
				Performance  &  87.71 / 88.74 / 98.63      & 92.73 / 92.41 / 99.44    & 91.41 / 90.87 / 99.89     &95.53 / 95.05 / 100   &90.06 / 89.47 / 99.76       \\ \hline
		\end{tabular}}
		\label{tab:general}
	\end{table*}

	\begin{figure*}[] 
		\centering 
		\includegraphics[width=0.85\textwidth]{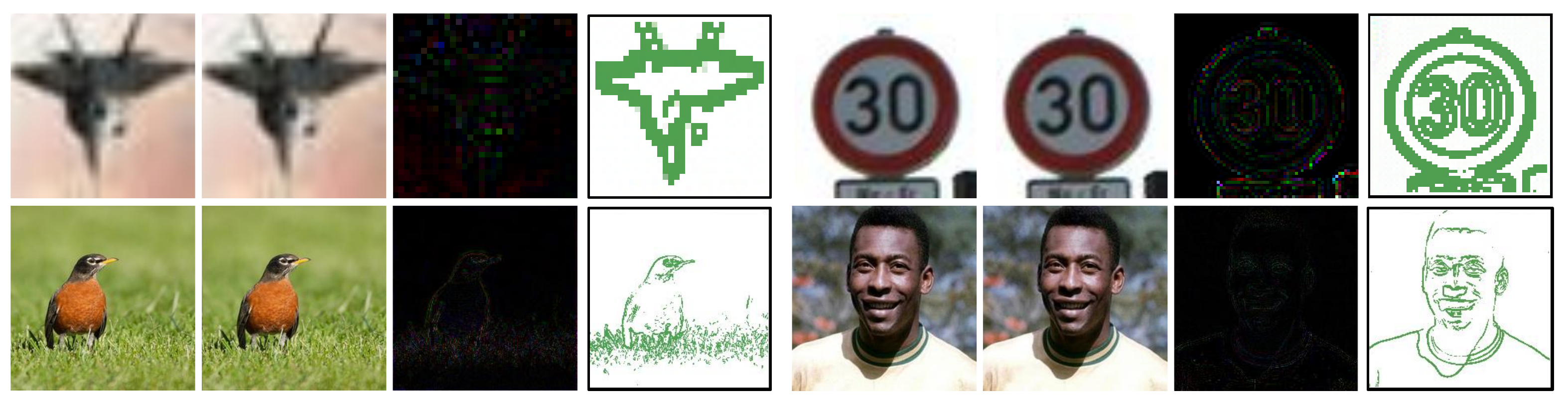} 
		\figcaption{Visual examples of Poison Ink on different datasets. For each dataset, the three images denote clean image, poisoned image, their 10$\times$ difference, \zj{and the original edge trigger extracted by $\mathbf{\mathit{GE}}$, respectively.} } 
		\label{fig:4vis} 
	\end{figure*}

	\begin{table}[]
		\centering
		\footnotesize
		\tabcaption{Quantitative results of the stealthiness of our Poison Ink on different datasets.} 
		\vspace{0.5em}
		\setlength{\tabcolsep}{5.5mm}{ 
			\begin{tabular}{c|c|c|c}
				\hline
				Invisibility  & PSNR $\uparrow$  & SSIM $\uparrow$  & LPIPS $\downarrow$  \\ \hline 
				CIFAR-10    & 42.95  & 0.9961  & 0.0001      \\ \hline
				GTSRB    & 38.05  & 0.9965  & 0.0001      \\ \hline
				ImageNet    & 41.62  & 0.9915  & 0.0020      \\ \hline
				VGG-Face & 45.34  & 0.9916  & 0.0011 \\ \hline
		\end{tabular}}
		\label{tab:vis_2} 
	\end{table}

	\subsection{Generality and  Flexibility of Poison Ink}
	To demonstrate the generality of our attack to different datasets and network structures, we run the controlled experiments with VGG-19 on different datasets and different network structures on CIFAR-10 dataset.  Poison Ink is imperceptible among various datasets, with visual and quantitative results shown in \Fref{fig:4vis} and \Tref{tab:vis_2}, respectively. In \zj{\Fref{fig:4vis}, we can see that the Guidance Extractor informs the Injection Network to preserve the edge trigger well. } Besides,  \Tref{tab:general} shows that our attack still guarantees a high ASR on different datasets and architectures while only decreasing the pristine performance slightly.

\begin{figure*}[]
		\centering
		\subfloat{\includegraphics[width=0.25\linewidth]{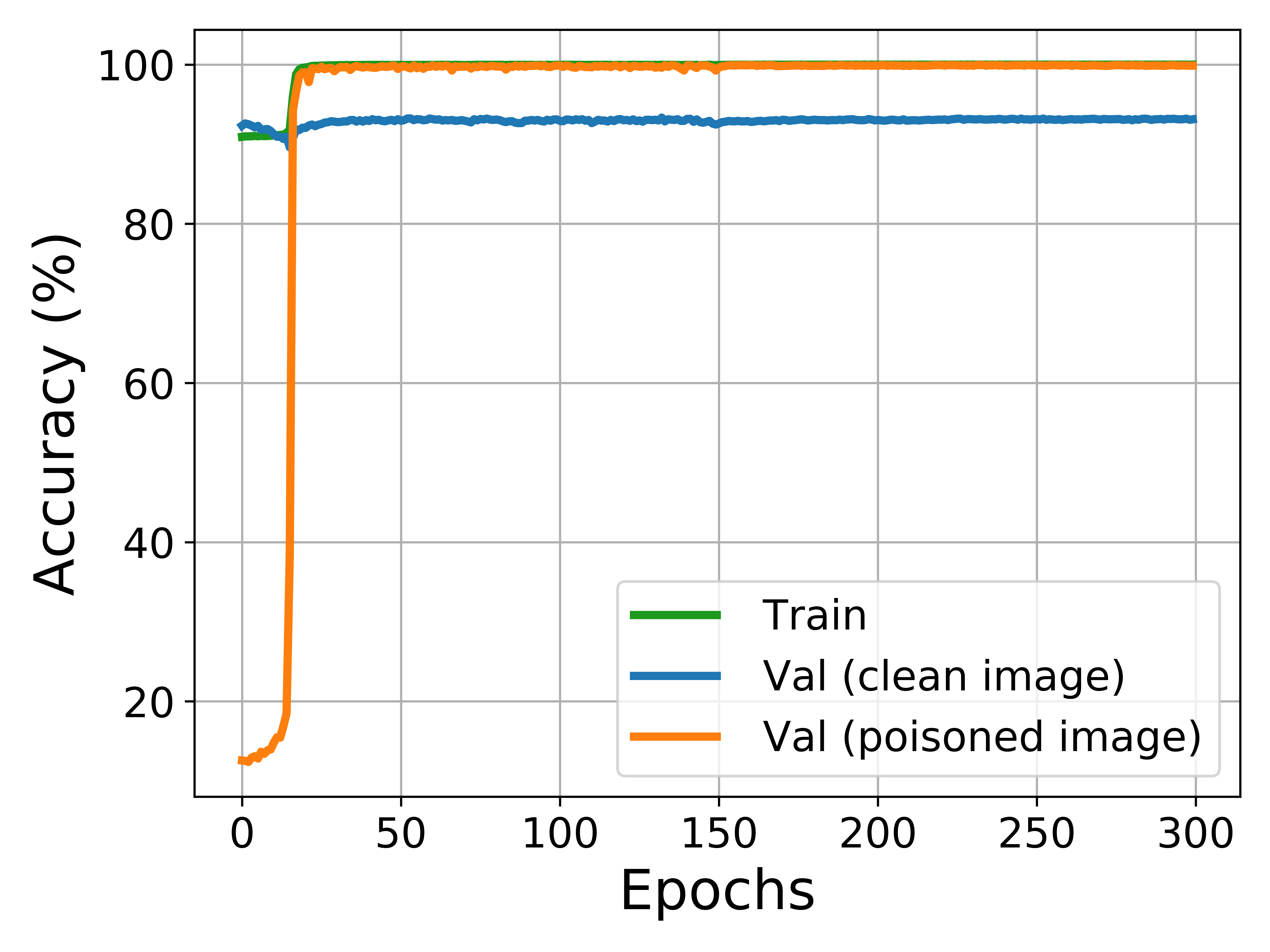}  \label{fig:pre}}
		\centering
		\hspace{1em}
		\subfloat{\includegraphics[width=0.25\linewidth]{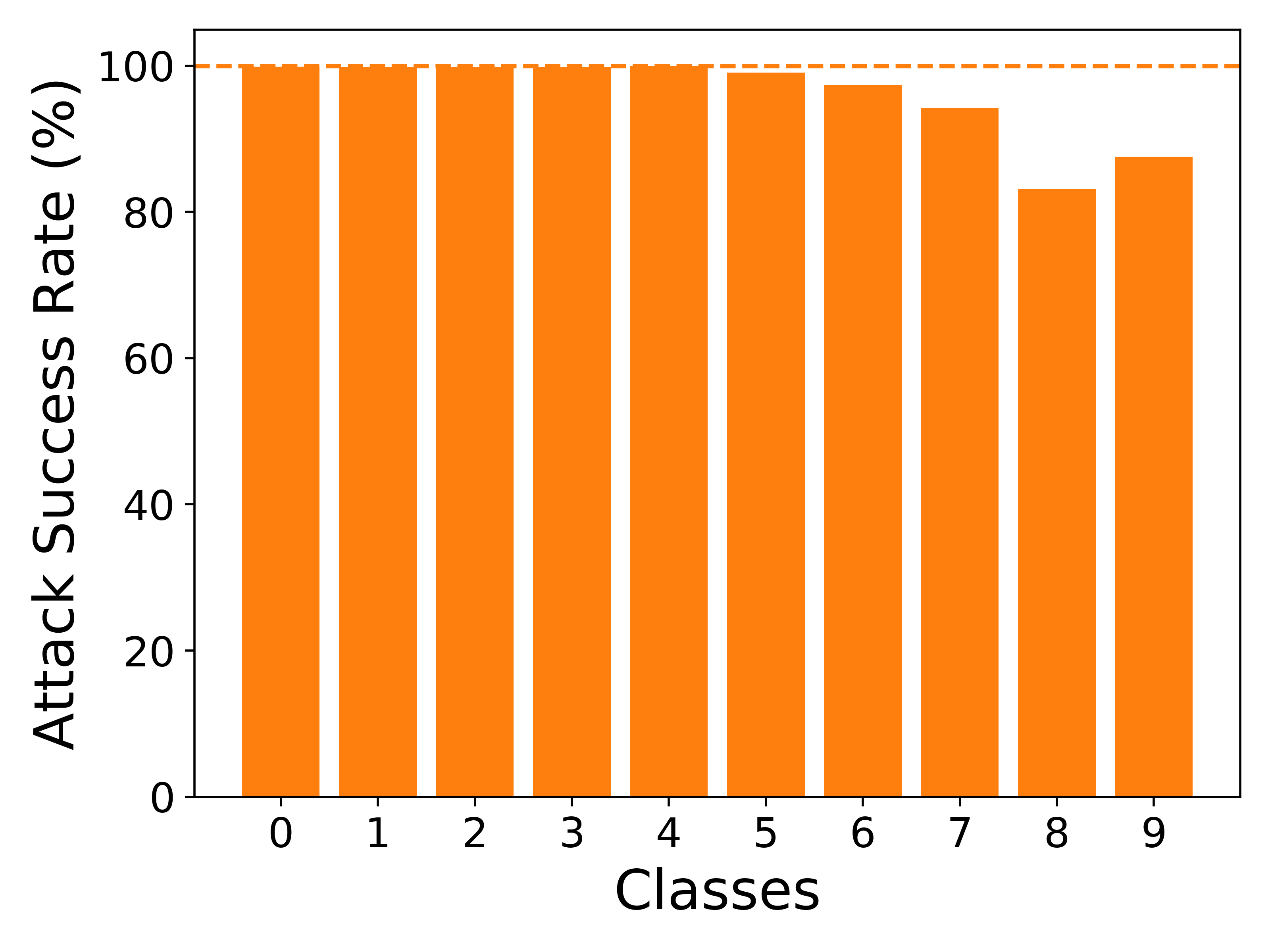} \label{fig:mul}}
		\hspace{1em}
		\subfloat{\includegraphics[width=0.25\linewidth]{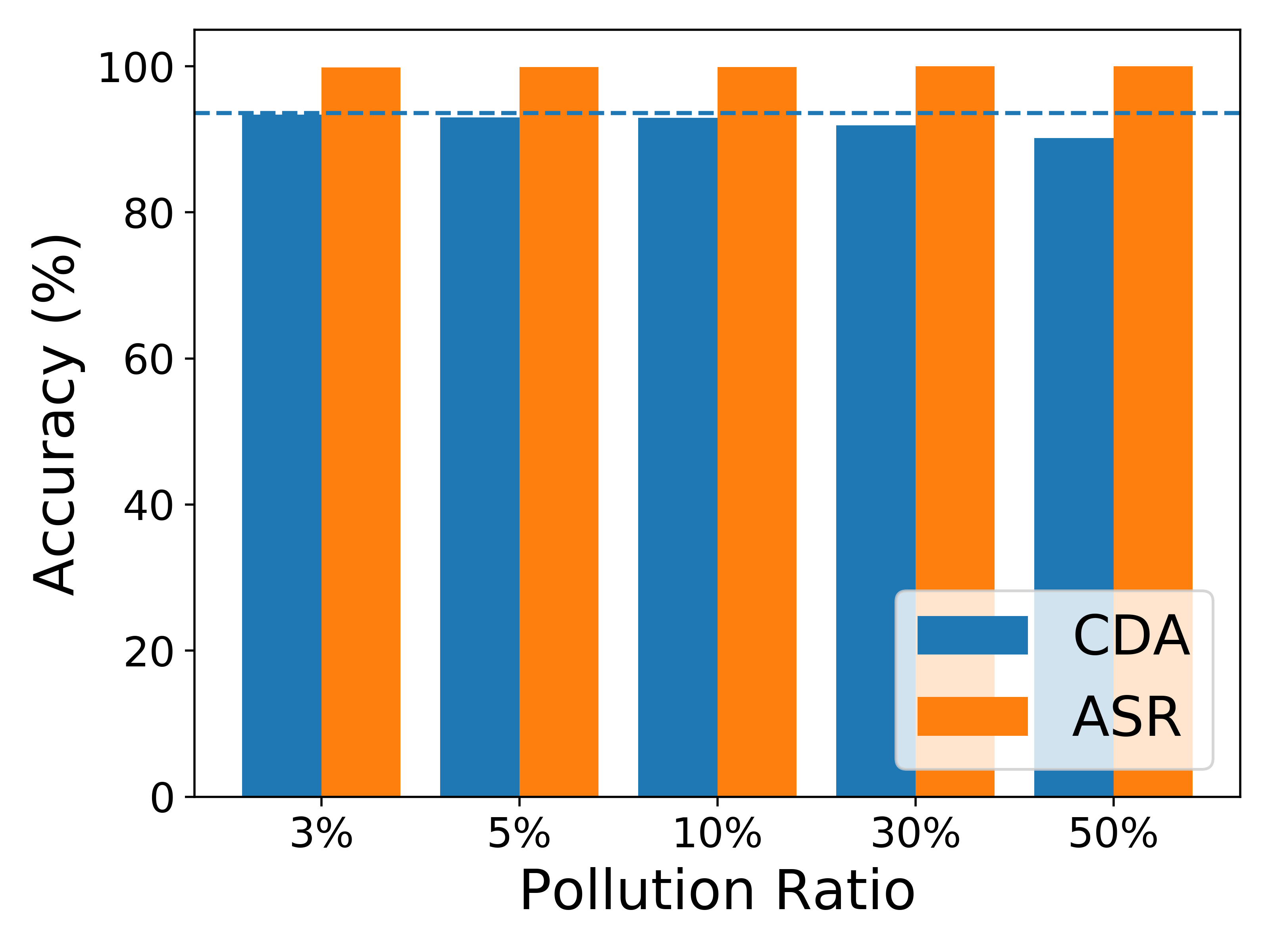} \label{fig:ratio}}
		\caption{The flexibility of Poison Ink. \textbf{Left:} the train-val convergence curve for training from pre-trained VGG-19; \textbf{Middle:} the attack success rate (ASR) of Poison Ink trained with multiple target labels, the orange dot line denotes the ASR of Poison Ink trained with single target label; \textbf{Right:} The performance (CDA/ASR (\%)) of infected model with different pollution ratios The blue dot line denotes the performance of original clean model. All three experiments are conducted on the CIFAR-10 dataset. }
		\label{fig:flex}
	\end{figure*}

	For the flexibility of Poison Ink, we further consider three more attack scenarios: training the target model from the pre-trained model, with multiple target labels, and under different ratios. All these experiments are conducted on CIFAR-10 dataset.
	In the former scenario, we set the initial learning rate as 0.001, and we observe in the left of \Fref{fig:flex} that the ASR achieves nearly 100\% after only 16 epochs.
	For the multiple-label attack, we inject 10 different poison ink (color) to generate the corresponding poisoned image for 10 target labels of CIFAR-10 dataset. As displayed in the middle of \Fref{fig:flex}, ASR for most target label attacks is comparable with single label attack, and so is the CDA (single:93.17, multiple:91.91).
	Besides, different ratios are also considered, and we can see on the right of \Fref{fig:flex} that Poison Ink attacks successfully under different ratios, which will be further discussed in the ablation study.

	\begin{figure}[!t]
		\centering
		\includegraphics[width=0.65\linewidth]{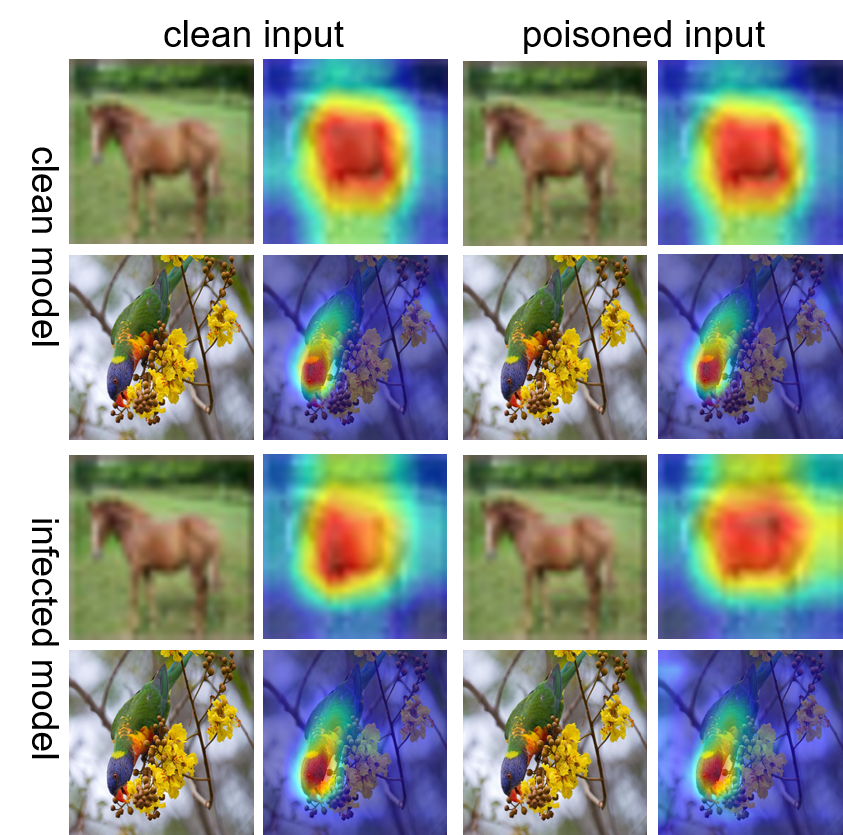} 
		\caption{ The Grad-CAM of clean input and poisoned input according to clean model and infected model. As shown in the figure,  Grad-CAM fails to detect trigger regions of those generated by our attack, which is indistinguishable with the benign case.}
		\label{fig:februus}
	\end{figure} 
	
	\begin{figure}[!t]
		\centering
		\includegraphics[width=1\linewidth]{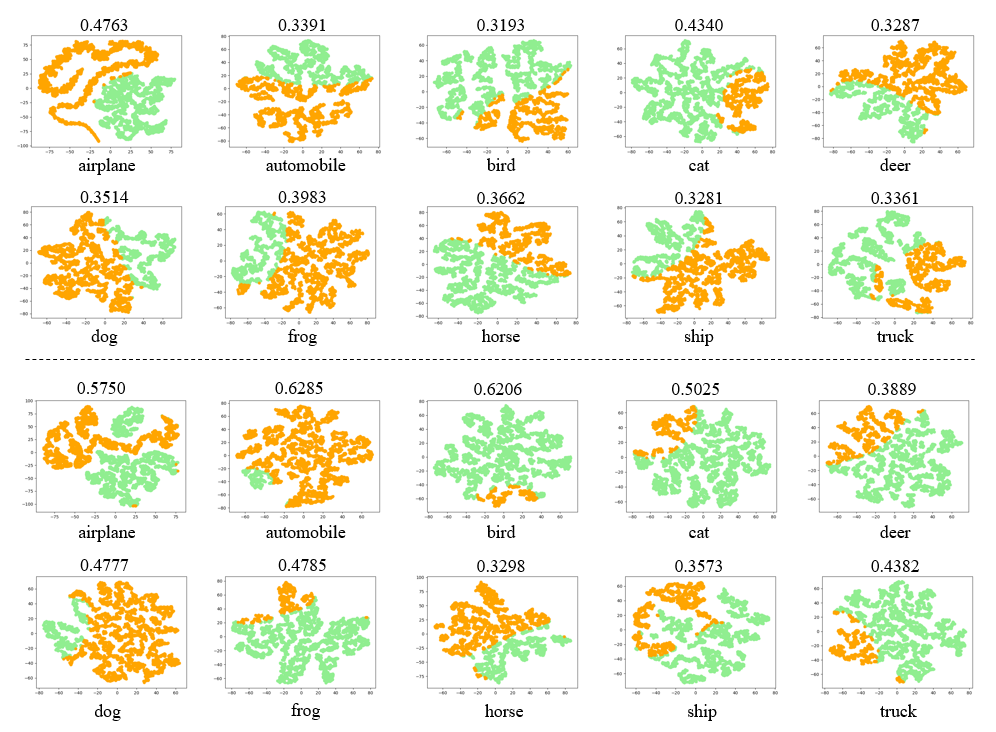} 
		\caption{ \zj{The activation clustering results for each class on the CIFAR-10 dataset. Silhouette score is given on the top of each subfigure, and a high silhouette score indicates that two clusters fit the data well. The top part is for the single label attack while the bottom part is for the multiple label attack.}}
		\label{fig:ac1}
	\end{figure} 
	
	\begin{figure*}[t]
		\vspace{-1em}
		\centering
		\subfloat[\footnotesize \textit{STRIP} \cite{gao2019strip}]{\includegraphics[width=0.48\linewidth]{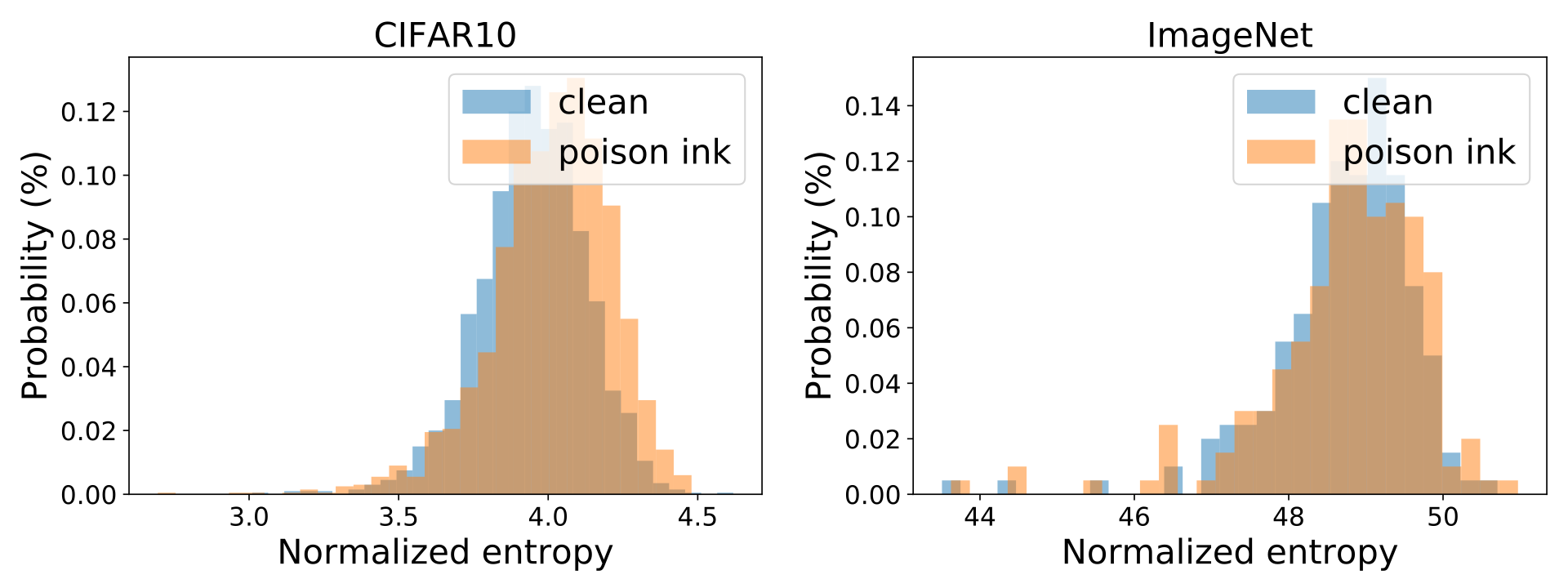}\label{fig:strip}}
		\subfloat[\footnotesize \textit{Fine-pruning} \cite{liu2018fine}]{\includegraphics[width=0.48\linewidth]{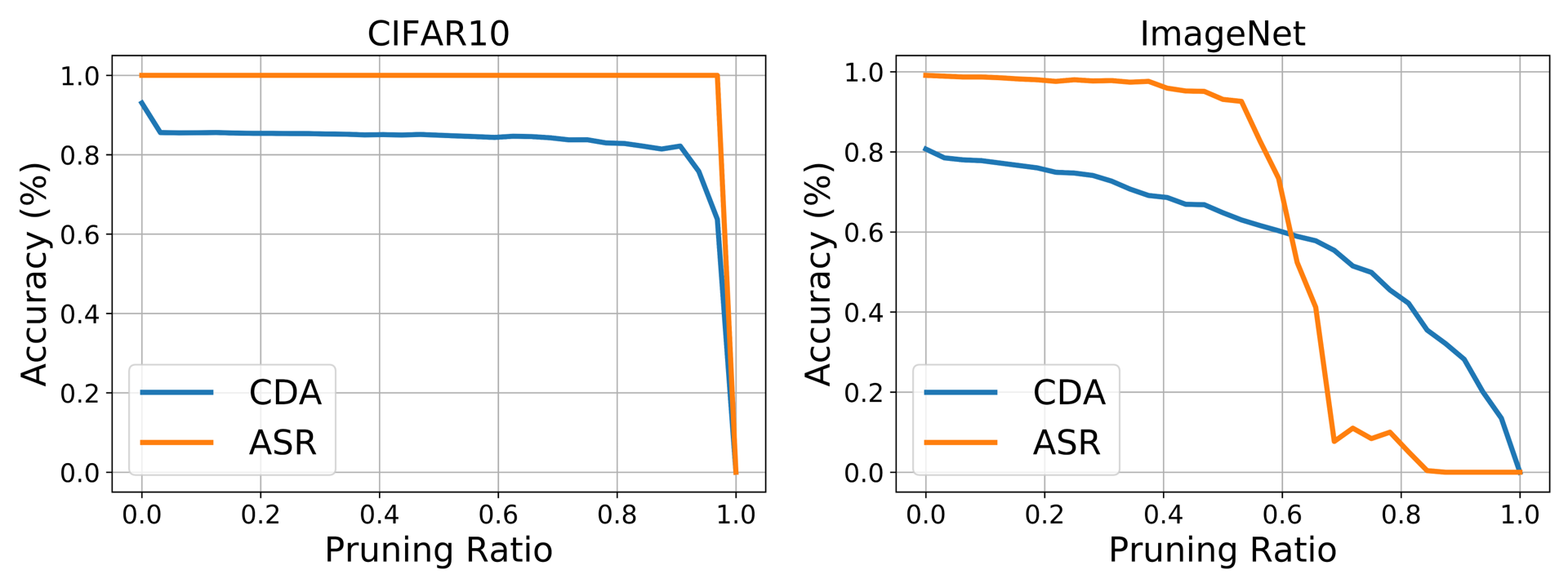}\label{fig:fp}}  \\
		\vspace{0.5em}
		\subfloat[\footnotesize \textit{Neural Cleanse} \cite{wang2019neural}]{\includegraphics[width=0.24\linewidth]{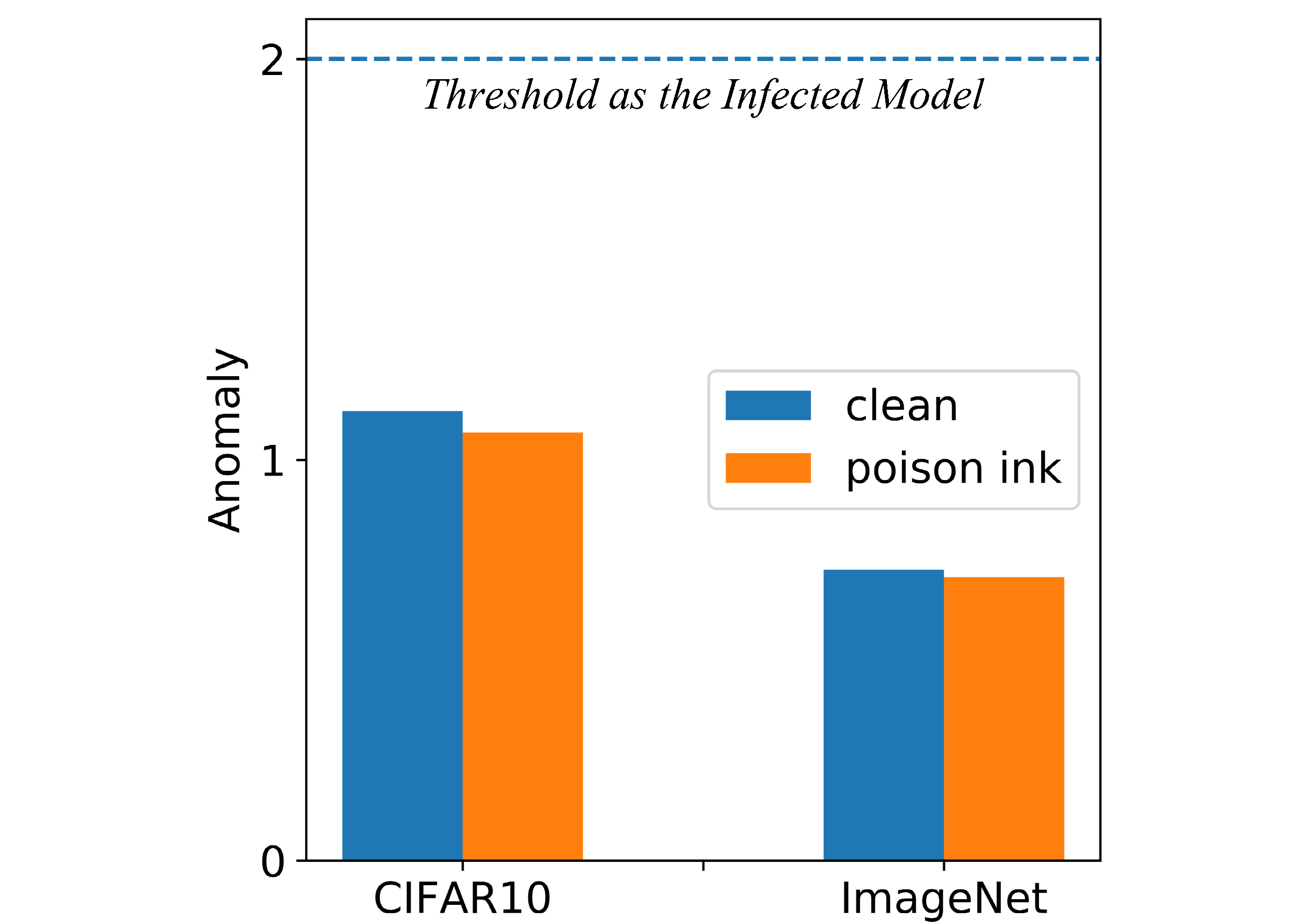} 	\label{fig:nc}} 
		\subfloat[\footnotesize \textit{TABOR} \cite{guo2019tabor}]{\includegraphics[width=0.25\linewidth]{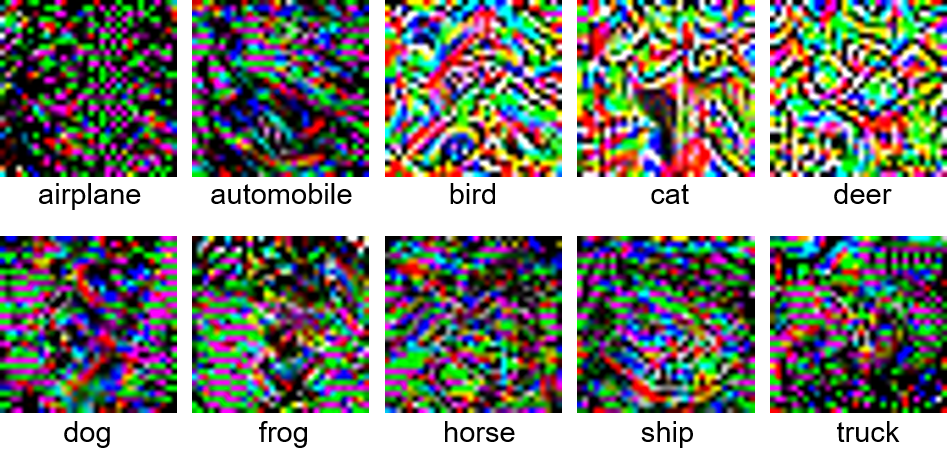} \label{fig:tabor}} 
		\hspace{1em}
		\subfloat[\footnotesize \textit{MESA} \cite{qiao2019defending} ]{\includegraphics[width=0.43\linewidth]{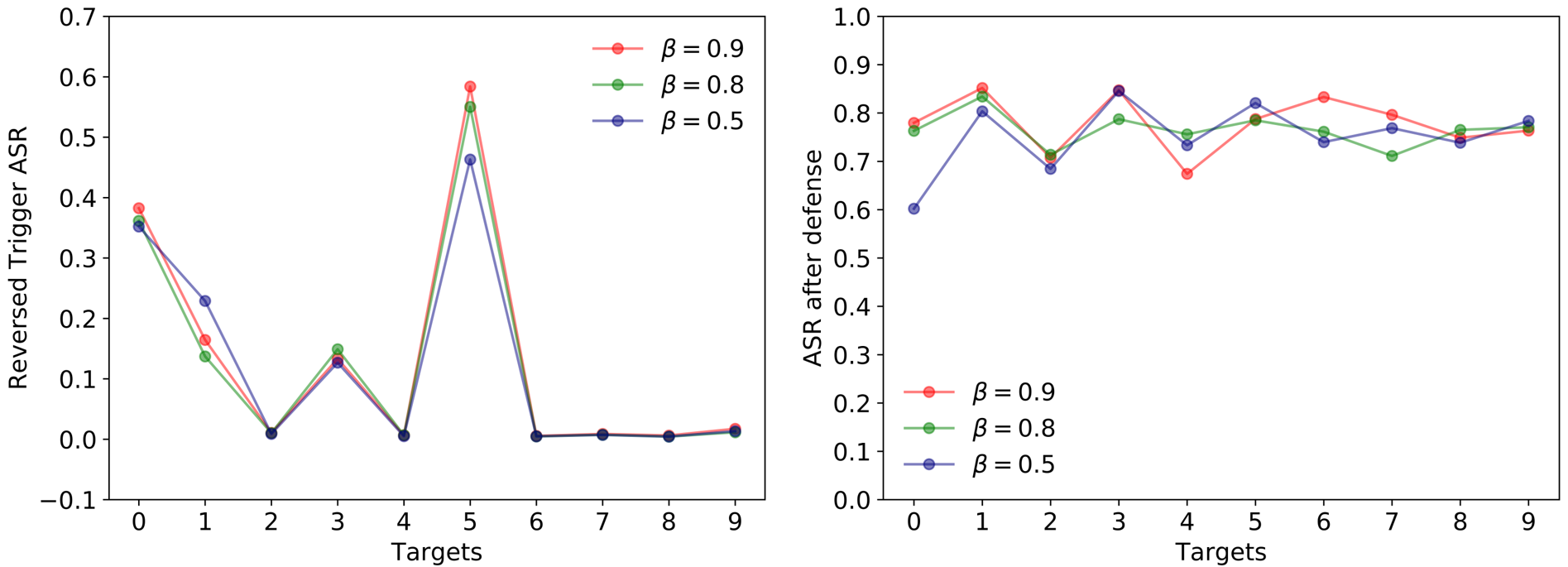}\label{fig:mesa}}  
		\caption{Some visual and quantitative results showing the resistance of Poison Ink to some state-of-the-art defense techniques.}
	\end{figure*}

	\subsection{Resistance to Defense Techniques}
	\label{sec:defense}
	To resist backdoor attacks, many different defense techniques have been proposed recently. In this section, we will test the resistance of Poison Ink against different defense techniques, which are further categorized as data-based defense, model-based defense, and defense with the meta classifier.
	
	\vspace{0.2em}
	\noindent \textbf{Data-based defense.}
	\textit{Februus} \cite{doan2019februus} utilizes Grad-CAM \cite{selvaraju2017grad} to visualize the attention map of the target image and regards the area with the highest score as
	the trigger region, then removes this region and restores it with image inpainting techniques. We first provide the attention map in \Fref{fig:februus}. During visualization, we feed the clean model and our infected model with both  clean images and the corresponding poisoned images generated by Poison Ink. We find that the infected model also focuses on the main content area of the input, which is similar to the clean model. Then, we remove such a region and directly restore it with the same region of the corresponding clean image. After trigger pattern removal, the ASR of Poison Ink still guarantees  98.84\% and 98.01\% on CIFAR-10 and ImageNet, respectively. 
	
	\textit{STRIP} \cite{gao2019strip} blends input images with a set of clean images from different classes and compares the entropy of the prediction before and after blending. 
	As shown in \Fref{fig:strip}, the entropy distribution of Poison Ink looks very similar to that of the clean model, which means that Poison Ink behaves normally as the benign (clean) model.
	
	\vspace{0.2em}
	\noindent \textbf{Model-based defense.}
	For model-based defense, we first try the totally white-box defense \textit{Fine-pruning} \cite{liu2018fine}. As shown in \Fref{fig:fp}, the ASR of Poison Ink remains 92.6\% with a 50\% pruning rate on the ImageNet, while the performance on clean data decreases significantly. 
	On CIFAR-10 dataset, we observe that the ASR of Poison Ink still stays at 100\% even with a 90\% pruning rate. 
	Next, we run the official code of ABS \cite{liu2019abs} on the only supported CIFAR-10 dataset, and find ABS fails to detect our method.
	\zj{In \Fref{fig:ac1}, we find that \textit{Activation Clustering} \cite{chen2018detecting} can infer the infected class (``airplane'') by a larger silhouette score, which is based on the assumption that the infected class will induce more overfitting. However, it cannot find the true poisoned images with a large false alarm (TPR:0.0002, FPR:0.984). Moreover, poison ink can pollute all labels, while \textit{Activation Clustering} cannot handle this case, as shown in the bottom of  \Fref{fig:ac1}.}
	
	Then, we consider  \textit{Neural Cleanse} \cite{wang2019neural}, which is designed for small and static triggers.
	It first reverses the trigger pattern based on the optimization method, then runs an anomaly detection among all reversed trigger patterns from each label. It defines a model as the infected model by Anomaly Index larger than the threshold $t = 2$. As shown in \Fref{fig:nc}, the Anomaly Index for our backdoored model is both below the threshold on CIFAR-10 dataset and ImageNet dataset. Besides, the infected label found by \textit{Neural Cleanse } is empty on CIFAR-10 dataset, and is wrong on ImageNet Dataset (\#97 not \#0). 
	Based on \textit{Neural Cleanse}, \textit{TABOR} \cite{guo2019tabor} adds more constraints for the reversed trigger pattern during the optimization process, which reduces the false positive rate but costs more computation time. The corresponding Anomaly Index on CIFAR-10 dataset is 1.5938 \zj{(clean mode:1.0806)}, and some visual results of reversed triggers are further shown in \Fref{fig:tabor}. On ImageNet, \textit{TABOR} costs 7 hours for a single label, and there still exist false alarms of the Anomaly Index (clean model:4.1593; Poison Ink:2.3125). 
	
	Rather than using the single reversed trigger, MESA \cite{qiao2019defending} models the trigger distribution and uses many generated triggers to implement backdoor removal. Once our backdoored model attacks all the labels, it can only find partially polluted labels, and our method can still keep a high ASR after backdoor removal, as shown in \Fref{fig:mesa}.
	For \textit{TND}, it provides a faster detection in the data-scarce scenario, but it can also be evaded by Poison Ink easily.

	\vspace{0.2em}
	\noindent \textbf{Meta classifiers.}
	We further try the recent meta classifier based defense method \textit{ULPs} \cite{kolouri2020universal} on CIFAR-10 dataset as an example. \textit{ULPs} classifies the suspect model as ``clean" or ``corrupted" by feeding the universal patterns to the suspect model and analyzing its output. The fewer the universal patterns required, the weaker the backdoor attack is. \zj{In general, 5 Universal Litmus Patterns (ULPs) can detect backdoor attacks successfully. The corresponding experiment shows that our backdoored model is classified as the clean one even with 10 ULPs. }
	
	\subsection{Adaptive Robustness Evaluation}
	\zj{Adaptive robustness evaluation has become the standard in adversarial machine learning in recent years \cite{tramer2020adaptive,carlini2019evaluating}. In this section, we evaluate the resistance of Poison Ink against some adaptive defense techniques, which takes into account the fact that trigger patterns are not static and may be hidden in the image edges.}
	
	\zj{We first test the robustness of Poison Ink to adaptive pre-processing operations such as Gaussian Noise and Gaussian Blur, which may mitigate the Poison Ink by disrupting the image edges. In addition, we adopt three data augmentations used in \cite{borgnia2021strong} for sanitizing backdoor attacks. In detail, Mixup means blending inputs with clean images, while Cutout and CutMix denote that patches of inputs are randomly cut and pasted with nothing or clean images, respectively. We display quantitative results on ImageNet dataset and CIFAR-10 dataset in  \Tref{tab:gaussian}. The results suggest that Poison Ink can still keep a relatively high ASR after suffering from such pre-processing operations.}

	\begin{table*}[t]
		\footnotesize
		\centering
		\caption{Robustness of Poison Ink to other adaptive pre-processing operations. The variance of Gaussian Noise is set as 0.01, the kernel size of Gaussian Blur is 3$\times$3, and three pre-processing operations used in \cite{borgnia2021strong} are also adopted. The performance degradation is shown in the bracket.}
		\setlength{\tabcolsep}{9.2mm}{ 
			\begin{tabular}{c|c|c|c|c}
				\hline
				Dataset & \multicolumn{2}{c|}{ImageNet} & \multicolumn{2}{c}{CIFAR-10} \\ \hline
				Pre-processing               & CDA  (\%)                 & ASR  (\%)      &  CDA  (\%)                 & ASR  (\%)  \\ \hline 
				Gaussian Noise & \reshl{56.28}{-}{24.28}  & \reshl{88.36}{-}{10.12}   & \reshl{55.17}{-}{37.75}  & \reshl{89.49}{-}{10.43} \\  \hline
				Gaussian Blur  & \reshl{73.84}{-}{6.72}  & \reshl{90.36}{-}{8.12} &  \reshl{52.53}{-}{40.39}  & \reshl{40.60}{-}{59.32} \\  \hline
				Mixup                      & \reshl{22.78}{-}{57.78}   & \reshl{73.62}{-}{24.86}  &  \reshl{45.77}{-}{47.15}  & \reshl{85.74}{-}{14.18}  \\  \hline
				Cutout                     & \reshl{56.44}{-}{24.12}  & \reshl{90.40}{-}{8.08} &  \reshl{65.82}{-}{27.10}  & \reshl{72.39}{-}{27.53} \\  \hline
				CutMix                   & \reshl{47.24}{-}{33.32}   & \reshl{90.62}{-}{7.86}  &  \reshl{61.79}{-}{31.13}  & \reshl{85.11}{-}{14.11}  \\  \hline
		\end{tabular}}
		\label{tab:gaussian}
	\end{table*}
	
\begin{figure}[!t]
		\centering
		\includegraphics[width=0.9\linewidth]{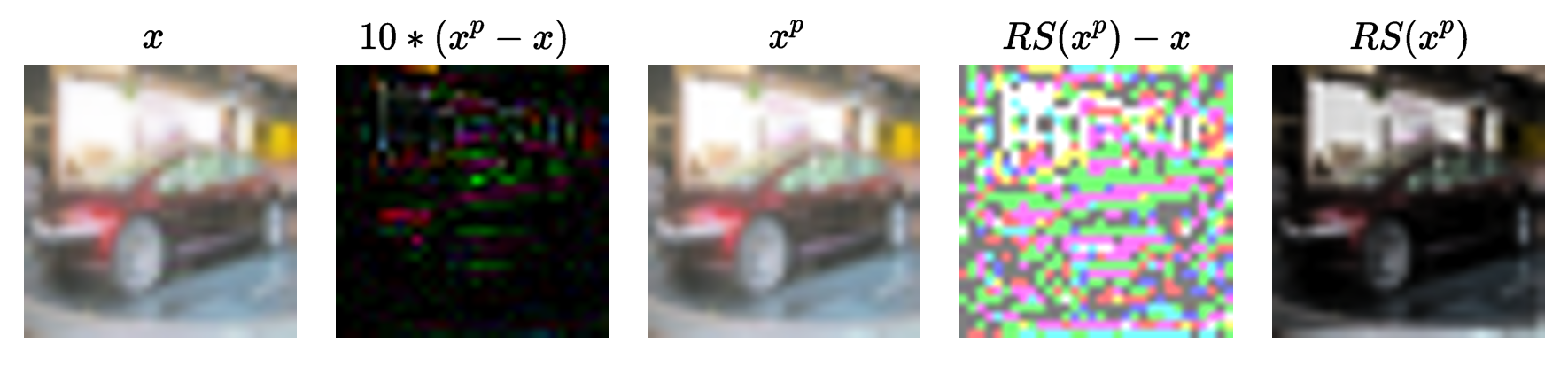} 
		\caption{ Visual examples of inputs processed by RCA-SOC (RS)\cite{chen2020rca}.  }
		\label{fig:rs}
	\end{figure}

	\zj{RCA-SOC \cite{chen2020rca} is proposed as a defense against evasion attacks by refocusing on critical areas and strengthening object contours. Specifically, pixel channel attention is adopted to focus on the critical feature areas, and pixel plane attention is designed to focus more on feature pixels, where the key pixels of the image are emphasized and the adversarial perturbed pixels are weakened.
	We utilize RCA-SOC to filter the inputs from CIFAR-10 dataset and find the CDA degrades heavily (from 92.92\% to 11.47\%) while the ASR (77.75\%) is still acceptable. We further showcase some visual examples in \Fref{fig:rs}, and we find that the perturbed pixels by Poison Ink are preserved to some extent after RCA-SOC, which makes our attack still succeed.}
	
	\begin{figure}[!t]
		\centering
		\includegraphics[width=1\linewidth]{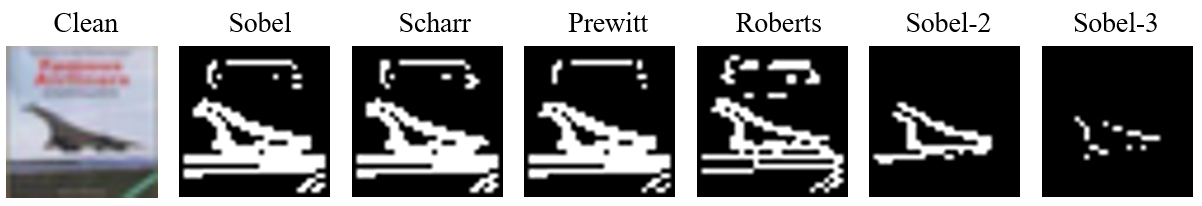} 
		\caption{ Visual examples of edge structures extracted by different edge extraction algorithms.  }
		\label{fig:edge}
	\end{figure} 
	
	\begin{table}[t]
		\footnotesize
		\centering
		\caption{\zj{The performance (CDA / ASR) in the case of replacing edge areas with the corresponding original (ori) clean image or the pure image with a constant RGB value. We utilize different edge  extraction algorithms to locate the edge area.}}
		\setlength{\tabcolsep}{1.8mm}{
			\begin{tabular}{c|c|c|c|c|c|c}
				\hline
				(\%)     & Sobel & Ccharr & Prewitt & Roberts & Sobel-2 & Sobel-3 \\ \hline
				CDA\_pure & 60.98 & 60.71  & 61.96   & 65.3    & 84.17  & 90.49  \\ \hline
				ASR\_ori & 66.39 & 65.57  & 67.76   & 74.57   & 94.56  & 99.14  \\ \hline
				ASR\_pure & 57.28 & 56.59  & 58.34   & 62.15   & 88.25  & 97.89  \\ \hline
		\end{tabular}}
		\label{tab:edge}
	\end{table}
	
	\zj{To remove the poison information more accurately, we further assume that the defender can utilize edge extraction algorithms to locate the edge area and then replace this area with another image. For edge extraction algorithms, we consider the same algorithm (``Sobel'') used for the trigger pattern generation,  other algorithms (``Scharr'',``Prewitt'',``Roberts'') with the same parameters, and the same algorithm with different parameters (``Sobel-2'',``Sobel-3''). Some visual examples are provided in \Fref{fig:edge}. For the replacement of the edge area, we first consider the above-mentioned solution, namely, directly using the corresponding original clean image (denote as ``ori''), which however is not accessible in practice. Nevertheless, the defender can leverage an inpainting algorithm to restore the removed area  \cite{doan2019februus}. Here, we use a pure image with a constant RGB value (\eg 125) as the restored area, which can be regarded as the worst case of inpainting. As shown in \Tref{tab:edge}, Poison Ink can still achieve above 55\% ASR in all cases mentioned above. We explain that the poison information is not fully removed in such adaptive defenses. Besides, in the case with a relatively low ASR, the corresponding CDA also degrades a lot, which is unacceptable for the defender. }
	
	\zj{We also consider adversarial training \cite{madry2017towards}, which is proposed to defend against adversarial attacks. In detail, Geiping \etal \cite{geiping2021doesn} extend adversarial training to defend against backdoor attacks by generating poisoned images during training and injecting them into training batches. We conduct the corresponding experiments on CIFAR-10 dataset, adopt $s$ = 0.75 as the poison immunity, and consider the \textit{from-scratch} scenario. To generate poisoned images, we utilize the most similar watermark-based method \cite{shafahi2018poison} provided by the official code \footnote{\href{https://github.com/JonasGeiping/data-poisoning}{https://github.com/JonasGeiping/data-poisoning}}. However, the ASR of Poison Ink even increases by 6.49\%  after adversarial training. The possible reason is that adversarial training forces the model to capture more robust features, making the edge structures learned better.}
	\zjj{This finding warrants more study in future work, as adversarial training is one of the only truly effective methods we have to date to defend against adversarial attacks. }

	\begin{table*}[t]
		\centering
		\caption{Comparison with other attack methods under different pollution ratio. We take results on CIFAR-10 dataset for example.}  
		\setlength{\tabcolsep}{4.8mm}{ 
			\begin{tabular}{c|c|c|c|c|c|c|c|c}
				\hline
				Metric               & Ratio &  BadNets \cite{gu2019badnets}& Blend \cite{chen2017targeted} & SIG \cite{barni2019new} &  Refool \cite{liu2020reflection} & SPM \cite{zhong2020backdoor}  &LSB \cite{li2019invisible} &   Ours  \\ \hline
				\multirow{4}{*}{CDA} & 1\%  &88.20  & 89.55 & 89.71  & 89.72  & 89.58    & 89.26 & 89.25 \\ \cline{2-9} 
				& 3\%  & 87.38 & 89.89 & 89.74 & 89.20 & 88.98 & 88.18 & 89.65 \\ \cline{2-9} 
				& 5\%  & 87.13 & 89.60 & 89.64 & 89.16 & 88.90 & 86.98 & 89.69 \\ \cline{2-9} 
				& 10\% & 85.44 & 89.69 & 89.34 & 88.65 & 89.07 & 83.63 & 89.51 \\ \hline
				\multirow{4}{*}{ASR} & 1\%  & 61.76 & 84.79 & 98.79 &75.86    & 57.15  & 10.32   & 10.51  \\ \cline{2-9} 
				& 3\%  & 66.55 & 89.39 & 99.23 & 87.16 & 58.53 & 10.91 & 94.22 \\ \cline{2-9} 
				& 5\%  & 65.36 & 90.99 & 99.47 & 89.79 & 57.69 & 11.67 & 93.58 \\ \cline{2-9} 
				& 10\% & 67.80 & 93.06 & 99.43 & 92.79 & 57.37 & 15.76 & 93.48 \\ \hline
		\end{tabular}}
		\label{tab:ratio}
	\end{table*}
	
	\begin{table}[t]
		\centering
		\footnotesize
		\tabcaption{\zj{Quantitative results of the stealthiness with different loss constraints on the CIFAR-10 dataset.} }
		\vspace{0.5em}
		\setlength{\tabcolsep}{3.8mm}{ 
			\begin{tabular}{c|c|c|c}
				\hline
				Loss Constraints  & PSNR $\uparrow$  & SSIM $\uparrow$  & LPIPS $\downarrow$  \\ \hline 
				$	\mathcal{L}_{inv}$    & 41.24  & 0.9959  & 0.0002      \\ \hline
				$\mathcal{L}_{adv}$  & 8.18  & 0.1482 & 0.2870      \\ \hline
				$	\mathcal{L}_{inv}$ \& $\mathcal{L}_{adv}$   & 42.95  & 0.9961  & 0.0001      \\ \hline
				
		\end{tabular}}
		\label{tab:loss} 
	\end{table} 
	
	\begin{figure}[t]
		\centering
		\includegraphics[width=0.9\linewidth]{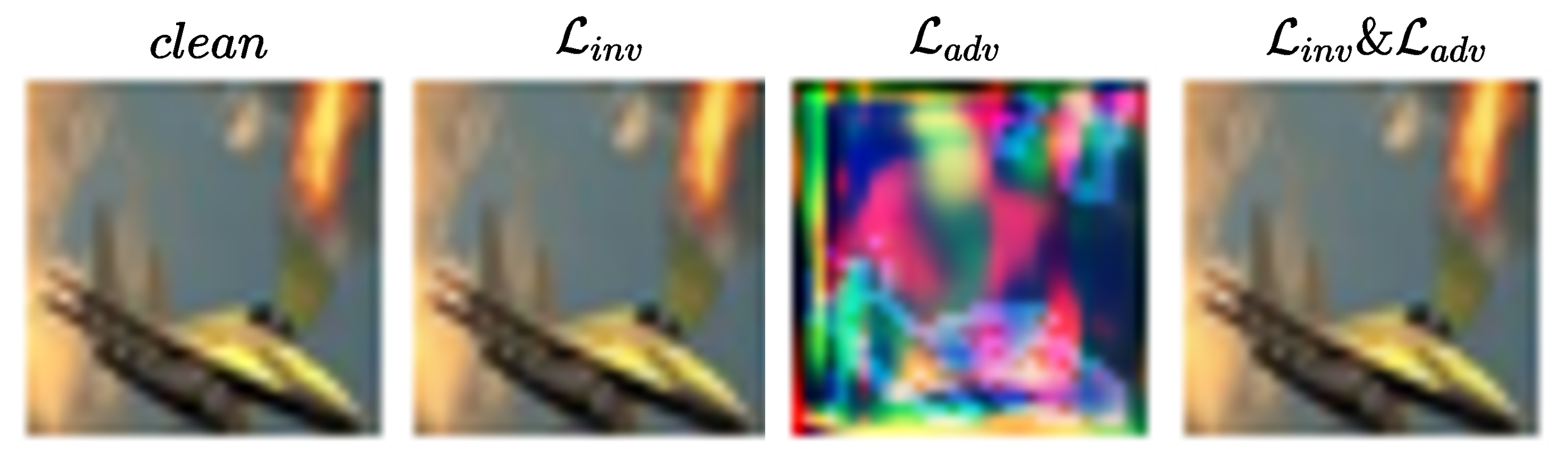} 
		\caption{\zj{Output examples of injection network $\mathbf{\mathit{IN}}$ trained with different loss constraints on CIFAR-10 dataset.}}
		\label{fig:loss}
	\end{figure}

	\subsection{Ablation Study} 
	\noindent\textbf{Importance of our design.} 
	In this experiment, we take CIFAR-10 dataset as an example to ablate our design. First, we use an input-agnostic image rather than the edge structure as the trigger pattern and leverage the same invisible injection network to create the poisoned image. However, we find this strategy will totally fail, and the ASR is only 13.42\% even without pre-processing. Very recently, Li \etal \cite{li2020backdoora} also proposed an invisible input-specific backdoor attack, whose framework is similar to us but does not considering the structure information. To double confirm it, we also conduct comparative experiments on CIFAR-10 dataset and find it fails as expected. 
	
	Second, we try to discard the interference layer during the injection network training. Under this setting, the ASR degrades from 99.95\% to 43.73\% when faced with resizing, demonstrating the importance of the interference layer.

	\vspace{0.2em}
	\noindent\textbf{Influence of pollution ratio.} In our default setting, we set the pollution ratio as 10\%. In \Tref{tab:ratio}, we further try more pollution ratios on CIFAR-10 dataset and show the  robustness and the stealthiness of different methods, where we use the average CDA and ASR to represent the robustness against data transformations. There exists a trade-off between the CAD and ASR under different ratios. Overall, compared with other methods, our method performs well both in stealthiness and robustness in most cases. 
	
	\zj{However, poison ink will fail under a meager pollution ratio like 1\%. In our threat model, we cannot control the training strategy. We try more invisible backdoor attacks such as WaNet \cite{nguyen2021wanet} and FTrojan \cite{wang2021backdoor} under the 1\% pollution ratio, and find they also fail with a low ASR in our threat model (WaNet: 10.83\% and FTrojan: 10.92\%). If we control the training strategy like in WaNet \cite{nguyen2021wanet}, namely, adding poisoned images in every training batch, Poison Ink can succeed with above 90\% ASR. Besides, if we first train the model only on trigger images and then fine-tune it on clean images, we can also achieve above 90\% ASR. In conclusion, the model needs more information to remember the trigger pattern when it is stealthy enough.}
	

	\vspace{0.2em}
	\noindent\textbf{Influence of Loss Constraints on Stealthiness.} \zj{In our default setting, we utilize both invisibility loss 	$	\mathcal{L}_{inv}$  and adversarial loss $\mathcal{L}_{adv}$  to achieve desirable stealthiness. As shown in \Tref{tab:loss}, appending adversarial loss $\mathcal{L}_{adv}$  after invisibility loss	$\mathcal{L}_{inv}$  can improve the image quality slightly, and only using $\mathcal{L}_{adv}$  will cause poor image quality, where injection network mainly focuses on the high-level information of images. Some visual results are also showcased in \Fref{fig:loss}.}
	\zjj{In practice, adopting or discarding the adversarial loss is flexible, which depends on the desired balance between stealthiness and robustness.}

	\section{Conclusion}
	In this paper, we point out the limitations of existing backdoor attacks regarding stealthiness and robustness. To address such limitations, we propose a new backdoor attack method ``Poison Ink". It utilizes the image structure as the carrier of poison information to generate trigger patterns and leverage a deep injection network to hide the trigger patterns into the cover images in an invisible way. Extensive experiments demonstrate that Poison Ink is superior to existing methods in stealthiness, robustness, generality and flexibility. 
	Besides, Poison Ink is resistant to many state-of-the-art defense techniques. It is interesting to explore backdoor attacks in the frequency domain, and we leave it as future work. 
	
	\vspace{1em}
	\noindent\textbf{Acknowledgement.} This work was supported in part by the Natural Science Foundation of China under Grant  U20B2047, 62072421, 62002334, 62102386 and 62121002, Exploration Fund Project of USTC under Grant YD3480002001, and by Fundamental Research Funds for the Central Universities under Grant WK2100000011 and WK5290000001. Gang Hua is partially supported by National Key R\&D Program of China Grant 2018AAA0101400 and NSFC Grant 61629301.

	{
		\bibliographystyle{IEEEtran}
		\bibliography{references}
	}

\end{document}